\documentclass[useAMS,usenatbib]{mn2e}

\usepackage{epsfig}
\usepackage{amsmath}

\newcommand{\bu}{\mbox{\boldmath $u$}}

\newcommand{\be}{\mbox{\boldmath $e$}}
\newcommand{\br}{\mbox{\boldmath $r$}}
\newcommand{\bB}{\mbox{\boldmath $B$}}
\newcommand{\bj}{\mbox{\boldmath $j$}}
\newcommand{\bO}{\mbox{\boldmath $\Omega$}}

\newcommand{\ptl}{\partial}
\newcommand{\dd}{{\rm d}}
\def\div{{\mathbf \nabla \cdot}}
\def\curl{{\mathbf \nabla \times}}
\def\grad{{\mathbf \nabla}}

\def\ephi{{ \hat{\be}_{\phi}}}

\def\rsun{R_{\odot}}

\def\Pr{{\rm Pr}}

\def\st{\sin\theta}
\def\ct{\cos\theta}
\def\s2t{\sin^2\theta}
\def\c2t{\cos^2\theta}

\title[Dynamics of the solar tachocline]{Dynamics of the solar tachocline -- II: the stratified case}
\author[P. Garaud \& J.-D. Garaud]{P. Garaud$^{1}$\thanks{E-mail:
pgaraud@ams.ucsc.edu} \& J.-D. Garaud$^{2}$ \\
$^{1}$Department of Applied Mathematics and Statistics, Baskin School of Engineering, University of California Santa Cruz \\
$^{2}$Centre des Mat\'eriaux, Mines Paristech/CNRS UMR 7633, Evry, France}

\begin{document}

\date{}

\pagerange{\pageref{firstpage}--\pageref{lastpage}} \pubyear{2007}

\maketitle

\label{firstpage}

\begin{abstract}
We present a detailed numerical study of the Gough \& McIntyre 
model for the solar tachocline. This model explains the uniformity 
of the rotation profile observed in the bulk of the radiative zone 
by the presence of a large-scale primordial magnetic 
field confined below the tachocline 
by flows originating from within the convection zone. 
We attribute the failure of previous 
numerical attempts at reproducing even qualitatively Gough \& McIntyre's 
idea to the use of boundary conditions which inappropriately model
the radiative--convective interface. We emphasize 
the key role of flows downwelling from the convection zone in confining the 
assumed internal field. We carefully select the range of 
parameters used in the simulations to guarantee a faithful representation 
of the hierarchy of expected lengthscales. We then present, for the first
time, a fully nonlinear and self-consistent 
numerical solution of the Gough \& McIntyre model which 
qualitatively satisfies the following set of observational constraints:
(i) the quenching of the large-scale differential rotation below the tachocline
 -- including in the polar regions -- as seen by helioseismology (ii) 
the confinement of the large-scale meridional flows to the uppermost 
layers of the radiative zone as required by observed light element abundances and suggested by helioseismic sound-speed data.
\end{abstract}

\begin{keywords}
MHD -- Sun:magnetic fields -- Sun:interior -- Sun:rotation
\end{keywords}

\section{Introduction}

The presence of the tachocline, a thin shear layer located at the interface 
between the radiative and convective regions of the Sun, was established two 
decades ago (Christensen-Dalsgaard \& Schou, 1988; 
Kosovichev, 1988; Brown {\it et al.} 1989; Dziembowski {\it et al.} 1989) 
but its {\it modus operandi} still remains mysterious. 

Anisotropic turbulent stresses associated with rotationally 
constrained eddies are thought to maintain 
the differential rotation profile observed within the convection zone:
\begin{equation}
\Omega_{\rm cz}(\theta,r) \simeq \Omega_{\rm eq}(r) (1-a_2(r) \cos^2\theta 
- a_4(r) \cos^4\theta) \mbox{   ,   }
\label{eq:ocz}
\end{equation}
where for example at $r = 0.75\rsun$ $\Omega_{\rm eq}/2\pi  
= 463 {\rm nHz}$, $a_2 = 0.17$ and $a_4  = 0.08$ (Schou {\it et al.} 
1998; Gough 2007). However, as shown by Spiegel \& Zahn (1992) 
(SZ92 hereafter), 
the mere reduction in the amplitude of these stresses naturally 
expected to occur across the radiative--convective interface 
(at $r_{\rm cz} = 0.713 \rsun$) cannot explain the transition to 
near-uniform rotation below.
It was later argued by Gough \& McIntyre (1998) (GM98 hereafter) 
that only {\it long-range} stresses can explain the suppression of 
the rotational shear in the bulk of the radiative zone 
and in addition maintain the angular velocity of the interior (observed to be 
$\Omega_{\rm rz}/2\pi \simeq 430$nHz) close to that of the surface despite 
the global spin-down induced by magnetic-braking.

Two competing theories for these presumed long-range stresses have been 
investigated: purely hydrodynamic stresses in the form of
gravity waves (see the review by 
Zahn, 2007) and hydromagnetic stresses (see the review by 
Garaud, 2007). This paper focuses on the latter mechanism only, although the 
dynamical balance in the Sun could arguably involve a combination of the two. 

It has long been known that even a very small primordial magnetic field 
embedded in the 
radiative zone could in principle impose uniform rotation 
(Ferraro, 1937; Mestel, 1953; Mestel \& Weiss, 1987). Ferraro's 
isorotation law,
\begin{equation}
\bB \cdot \grad \Omega = 0 \mbox{   ,   }
\end{equation} 
valid in the limit of negligible dissipation and for steady-state, 
axisymmetric flows, is usually stated as {\it the angular velocity 
must be constant on magnetic field lines}. Thus in its simplest form, 
Ferraro's law predicts the possibility of uniform rotation for the 
radiative zone provided the magnetic field is entirely confined beneath the
radiative--convective interface (R\"udiger \& Kitchatinov 1997). 
On the other hand, any field line directly connected with 
the convection zone promotes the propagation of the rotational shear into
the radiative zone (MacGregor \& Charbonneau 1999), 
inducing what will be referred to from 
here on as a ``differentially rotating Ferraro state''. 
Hence, field confinement is the key to the existence of a tachocline.

GM98 were the first to address the question of how
the presumed primordial field could indeed be confined within 
the radiative zone, and proposed
that meridional flows driven by Coriolis forces in the convection zone 
and burrowing downward would interact nonlinearly with the underlying
magnetic field lines,  bending them towards the horizontal in the
tachocline region, thus effectively suppressing direct radial
Alfv\'enic transport between the  convection zone and the radiative
zone. Conveniently, the same meridional flows can also be held
responsible for mixing light elements such as  Li and Be between the
convection zone and their respective nuclear-burning  regions, 
reducing He settling (Elliott \& Gough, 1999) as well
as providing weak but sufficient angular-momentum  transport to
adjust the mean rotation rates of the convective and
radiative zones continuously throughout the spin-down phase. The seminal
boundary-layer  analysis of the dynamics of the tachocline presented
by GM98 validated the plausibility of this theory, although
many of their simplifying scaling assumptions remain to be confirmed
through the direct numerical solution of the governing equations.

This paper first briefly reviews existing work and then 
presents new results on the laminar tachocline dynamics according to GM98. 
We begin by discussing past attempts at implementing 
their model numerically in Section \ref{sec:prevfail}. 
In particular, we argue that the failure of previous numerical studies 
in reproducing field confinement can be explained by 
the selection of inappropriate boundary conditions. A new numerical 
algorithm is then 
presented in Section \ref{sec:model}, and used in Section \ref{sec:numexp} 
to revisit the idea proposed by GM98 with much more success. 
We discuss future prospects in Section \ref{sec:disconcl}.

\section{Discussion of previous numerical models in the light of 
Gough \& McIntyre's original idea}
\label{sec:prevfail}

GM98 argue that the radiative interior should be
divided into three dynamically distinct zones including (from the base of 
the convection zone downward) (1) a more-or-less magnetic-free region in 
thermal-wind balance, well-ventilated by meridional flows originating 
from the convection zone, which can be 
thought of as the bulk of the tachocline, (2) a very thin magnetic 
advection-diffusion layer where the tachocline flows and the underlying 
field interact nonlinearly to confine one another and (3) a magnetically 
dominated, near-uniformly rotating interior. 

The nonlinear nature and geometric complexity of the problem precludes 
analytical solutions, and all existing studies since the 
original work of GM98 have been numerical (see the review by Garaud, 2007). 
Among these, only two include all the nonlinear terms required 
(i.e. the advection terms 
in the momentum and the magnetic induction equations, and the Lorentz 
force in the momentum equation)
to represent the nonlinear magnetohydrodynamics of the tachocline 
``correctly'': Garaud (2002) -- hereafter 
Paper I -- and Brun \& Zahn (2006) -- hereafter BZ06. Surprisingly, neither 
have been able to find evidence for the kind of dynamical balance proposed by 
GM98, and the time is therefore ripe to take a step back and discuss why.

\subsection{``Failure'' of previous numerical models}

Paper I and BZ06 are two studies which model magnetohydrodynamic
perturbations induced in the solar radiative zone by 
the differentially rotating convection zone and by an assumed 
primordial magnetic field. Both papers focus on the 
issues of field confinement and the suppression of differential 
rotation below the tachocline. 

Paper I presents steady-state, axially symmetric solutions of 
an incompressible and
unstratified version of the GM98 model. BZ06 on the other
hand use a time-dependent, three-dimensional 
algorithm based on the ASH code (Glatzmaier 1984; Clune {\it et al.} 1999; 
Miesch {\it et al.} 2000; Brun {\it et al.} 2004), and solve 
the complete set of anelastic MHD equations. 

Both studies otherwise consider 
a similar computational domain, namely a spherical shell which 
spans the region between the base of the 
convection zone (at $r = r_{\rm out}\simeq r_{\rm cz}$) and 
an inner sphere (at $r = r_{\rm in}$). The boundary conditions are also 
essentially similar: the ``outer'' boundary, 
which effectively models the radiative--convective interface,
is in both cases assumed to be impermeable and rotating differentially with 
an angular velocity profile given by equation (\ref{eq:ocz}); the 
magnetic field is matched onto a potential field. 

In Paper I, the numerical solutions only show confinement of the 
magnetic field lines in the equatorial
regions for low enough diffusivities,  occasionally at mid-latitudes
for lower field strength, but never in the  polar regions 
(see Figure 11 of Paper I for example). 
It is commonly argued that the failure of Paper I to find 
fully-confined magnetic field solutions stems entirely from the simplified 
nature of the model equations (incompressible, unstratified): indeed, 
within these assumptions the meridional flows are generated only by
Ekman-Hartmann pumping on the boundaries, and their amplitude scales with
the diffusivities in a way which always maintains the magnetic Reynolds number 
below unity. As a result, these 
flows are unable to confine the magnetic field and the
differential rotation imposed at the upper boundary of the domain 
persists within a large part of the radiative
zone.

The three-dimensional, time-dependent solutions of BZ06 
naturally depend on the initial magnetic field 
configuration selected. Various cases with an initial field more-or-less
deeply confined are discussed. Against expectations, BZ06 find that 
regardless of the initial conditions, the field lines always
spread out and eventually overlap with the convection zone, permitting the 
propagation of the differential rotation into the radiative interior. 
Interestingly, the transient state of the system -- prior to any field line
connecting with the convection zone -- qualitatively 
looks in many ways similar to the 
well-known GM98 picture, although it is clearly not a 
steady state. The ``tachocline'' thus formed slowly narrows with time until 
field lines spread into the convection zone, at which point a 
differentially rotating Ferraro state is rapidly established. 
One is forced to conclude in one of three ways: (1) for unspecified
primordial reasons, the initial magnetic field was more deeply embedded in 
the radiative zone and the currently observed tachocline is merely a transient
phase; (2) the tachocline is indeed in a steady-state and some of the 
assumptions made in the way have been wrong; (3) the parameter regime
studied by BZ06 (where diffusion coefficients are artificially 
increased by many orders of magnitude to satisfy numerical constraints) 
does not appropriately reproduce the solar interior dynamics. 

One should be uncomfortable in selecting the first option, as
it would place very strong and unlikely constraints on the initial field 
conditions to provide just the right structure for today's tachocline. One 
could naively tend to favour the third option, but BZ06 {\it a priori} 
took care to select a range of parameters for which all expected boundary layer
thicknesses are small enough and in the same hierarchical order as those
of the model proposed by GM98. 
The plot thickens while we are left with the uneasy task of reconsidering
the key features of either the numerical models or of the 
GM98 model (both, perhaps). 

\subsection{The source of the problem}
\label{subsec:sourceprob}

At this point it is worth discussing one of the more delicate aspects
of the GM98 model, namely the exact mechanism by which
these pivotal meridional flows are thought to be generated. This point
has been  a source of confusion and debate, but is clearly
crucial to a better  understanding of the tachocline dynamics. In 
the original work of GM98, the principal clue to the nature of
the flows can be found in the sentence:  {\it ``Turbulent stresses in
the convection zone induce (through Coriolis  effects) a meridional
circulation, causing the gas from the convection zone  to burrow
downwards ...''}. The flows considered by GM98 do not
originate from within the tachocline and can therefore  not be
appropriately modelled  by a numerical scheme in which the
radiative--convective interface is assumed to be impermeable (as in Paper I 
and BZ06). While this conclusion seems obvious
in hindsight, the physics of the problem are actually rather subtle and 
deserve clarification. In what follows, we discuss the issue in more detail
and present a unified view of the results of previous works on the subject. 

Spiegel \& Zahn were the first to study the
dynamics of the newly-discovered tachocline (SZ92). They considered
a non-magnetic radiative zone only, and imposed a 
latitudinally-varying rotation profile at the radiative--convective interface. 
They performed two distinct calculations. The first looked at the 
time-dependent evolution of the angular velocity profile in the 
radiative zone under such forcing, assuming isotropic viscous stresses. 
The second sought steady-state ``tachocline'' solutions assuming anisotropic 
turbulent stresses. Since the latter did not specifically address 
the question of the meridional flows, we focus here on the results 
of the time-dependent calculation, which can be interpreted 
in the following way. The differential rotation imposed by the 
convection zone to the top of the radiative zone inevitably induces
some degree of shear along the rotation axis if the radiative zone 
is originally in a state of uniform rotation. This generates a thermal
gradient in the latitudinal direction by way of the thermal-wind equation 
(see GM98, equation (1) for example). Meridional flows are then 
required to balance this thermal gradient when the system is in 
thermal equilibrium.
These flows burrow into the radiative zone, advecting angular momentum 
thereby helping the propagation of the shear further down. 
The results of SZ92 imply that the system continues evolving in this 
fashion until the radiative zone has achieved complete thermal and 
dynamical equilibrium.

The characteristic amplitude of the time-dependent flows associated with this
thermo-dynamical relaxation process is, by way of the assumptions 
listed above, only dependent on the (evolving) local differential rotation, 
stratification and thermal conductivity. Their turnover 
timescale is calculated to be of the order of the local Eddington-Sweet
timescale, which is short in the initial relaxation stages and steadily 
increases as the system evolves. 
Crucially, this result was derived independently of any
boundary condition on the meridional flows. It is therefore correct 
to think of the induced transient flows as being driven by baroclinic stresses 
from {\it within} the tachocline rather than downwelling from 
the convection zone. In fact, they exist {\it even if} the 
radiative--convective interface is assumed to be 
impermeable. This fact is probably at the origin of the impermeable boundary 
conditions selected by Paper I and by BZ06, in spite the obvious contradiction 
with GM98's intent. 

However, it is vital to remember that this transient phase and its associated
flows both end once the system achieves thermal and dynamical equilibrium. 
The only flows remaining in the final steady-state 
are driven, within tiny boundary layers, by the thermo(-magneto)-viscous 
stresses required to match the bulk equilibrium 
solutions to the applied boundary conditions. 
Thus, as expected from any elliptic problem, the nature of the boundary 
conditions selected entirely controls the steady-state solutions. 
This is clearly illustrated by the work of Garaud 
\& Brummell (2008) (GB08 hereafter), which complements that of SZ92 by
calculating the spatial properties as well as characteristic 
amplitudes of {\it steady-state} meridional flows in the radiative zone, 
as induced by various kinds of forcing applied at the radiative--convective interface. 

GB08 showed that in the non-magnetic case, 
the induced steady-state flows can always be viewed -- at least in the linear 
sense, and for solar parameters -- 
as the sum of two ``modes'': a viscously-dominated 
``Ekman mode'', which 
very rapidly decays away from the interface (on an Ekman lengthscale), 
and a ``thermo-viscous mode'' which can essentially span the 
entire radiative zone when not hindered by a magnetic field. 
Since the Ekman mode decays too rapidly to have
any effect on the tachocline dynamics, the flows discussed by 
GM98, which are thought to ventilate the bulk of the tachocline,
should be identified with the thermo-viscous mode. This view
is perfectly consistent with GM98's analysis, since the 
thermo-viscous mode is in thermal-wind balance (GB08).

The respective amplitudes with which the Ekman and thermo-viscous 
modes are driven
depend on the nature, amplitude and spatial structure of the forcing.
GB08 showed that when the interface is 
impermeable, as assumed in Paper I and by BZ06, then the amplitude 
of the thermo-viscous mode is negligible for microscopic solar values 
of the diffusivities. As a result, the magnetic Reynolds number of these
tachocline flows is much smaller than unity, which straightforwardly 
explains why field confinement eluded these two previous studies.

When flows are pumped directly through the interface by stresses 
within the convection zone (i.e. the radial component of the flows 
is non-zero at the radiative--convective interface), GB08
find that the amplitude of the thermo-viscous mode can be much higher, 
in which case the tachocline flows may be expected to confine the field 
in a scenario qualitatively similar to the one proposed by GM98. 
The quantitative detailed analysis must be done numerically; 
this is the purpose of the present study.

\section{The numerical model}
\label{sec:model}

We now revisit the idea proposed by GM98 in the light
of the previous discussion. A new numerical algorithm has been constructed
specifically for this purpose. We first present, for the sake of completeness, 
a derivation of our model equations in Section \ref{subsec:numeqs}. 
Since our computational domain is limited 
to a spherical shell within the radiative zone, we then present and 
discuss in detail all the boundary conditions applied in Section 
\ref{subsec:numbcs}. The relevant non-dimensional parameters, and 
expected boundary layers of the problem are discussed in Section 
\ref{subsec:bls}. The numerical method, as well as other numerical constraints, 
are finally outlined in Section \ref{subsec:numalgo}.

\subsection{The governing equations}
\label{subsec:numeqs}

The Standard Solar Model (SSM hereafter) provides an accurate representation 
of the thermal structure and composition of the interior of a hypothetical 
spherically symmetric, non-rotating, non-magnetic Sun. The excellent
match between the SSM and helioseismic observations  
confirms that the likely internal dynamics of the Sun only induce weak 
deviations in the background state thermodynamical quantities. 
The large-scale magnetohydrodynamics 
of the solar interior, thought to be responsible for the maintenance of 
the peculiar observed rotation profile, can therefore be thought of 
as perturbations upon the SSM. 

The quasi-steady SSM equations (which are valid on timescales
much shorter than the nuclear burning timescale) reduce to the following
set of four equations within the solar radiative zone (but outside of the
nuclear burning core):
\begin{eqnarray}
&& - \grad \overline{p} = \overline{\rho} \grad \overline{\Phi} \mbox{   ,   } \nonumber  \\
&& \div( \overline{k} \grad \overline{T}) = 0   \mbox{   ,   }\nonumber  \\
&& \overline{p} = \overline{p}(\overline{\rho},\overline{T}) \mbox{   ,   } \nonumber  \\
&& \grad^2 \overline{\Phi} = 4\pi G \overline{\rho}  \mbox{   .   } 
\label{eq:backg} 
\end{eqnarray}
These equations describe hydrostatic equilibrium, thermal 
equilibrium, the equation of state and finally the Poisson equation for the 
gravitational potential respectively. Here, $\rho$, $T$, and $p$ 
are the standard
thermodynamical variables, $k$ is the thermal conductivity 
(which typically depends on $T$ and $\rho$), $G$ is the gravitational constant 
and $\Phi$ is the gravitational potential. The equation of state itself 
is well-approximated by that of a perfect gas in this region of the Sun. 
The solution to the set of equations (\ref{eq:backg}) for the present-day Sun
can be inferred from model S  of Christensen-Dalsgaard {\it et al.} (1991) for example 
(see Appendix A).

We then consider perturbations on this background equilibrium 
caused by meridional and azimuthal flows, as well as magnetic fields.  
For simplicity, we restrict our study to axisymmetric systems. Moreover, 
since the 
observed internal rotation profile of the Sun has remained approximately 
constant over the past decade, we boldly 
postulate that the Sun is in fact in a 
state of quasi-steady dynamical and thermal balance. 
The general set of equations governing the 
axisymmetric, quasi-steady perturbations are then 
\begin{eqnarray}
&& \rho \bu \cdot \grad \bu 
= - \grad p - \rho \grad \Phi + \bj \times \bB + \div \Pi \mbox{   ,   } \nonumber \\
&&  \div(\rho \bu) = 0 \mbox{   ,   } \nonumber  \\
&&  \rho T \bu\cdot \grad s  
= \div( k \grad T)\mbox{   ,   } \nonumber  \\
&& p = p(\rho,T) \mbox{   ,   }\nonumber \\
&& \grad^2 \Phi = 4\pi G \rho \mbox{   ,   } \nonumber \\
&& \curl(\bu \times \bB) = \curl( \eta \curl \bB)\mbox{   ,   } 
\nonumber  \\
&& \div \bB = 0\mbox{   ,   }
\label{eq:global}
\end{eqnarray}
where $\bu$ represents the axisymmetric flow velocity, 
$\bB$ the magnetic field and $\bj = \curl \bB / 4\pi $ is the electric current, $\Pi$ is the 
viscous stress tensor, $s$ is the specific entropy
and $\eta$ is the magnetic diffusivity. 
These equations respectively characterise the conservation of
 momentum, mass, and thermal
energy, the equation of state and the Poisson equation, conservation of 
magnetic flux and finally, the solenoidal condition. 

Thermodynamical perturbations are from here on denoted with tildes, 
as for example in $p = \overline{p} 
+ \tilde{p}$. The system of equations (\ref{eq:global}) can be 
linearised in the thermodynamical quantities 
around the background state equations (\ref{eq:backg}) to yield
\begin{eqnarray}
&& \overline{\rho} \bu \cdot \grad \bu 
= - \grad \tilde{p} - \overline{\rho} \grad \tilde{\Phi}  - \tilde{\rho} \grad \overline{\Phi}  + \bj \times \bB + \div \Pi \mbox{   ,   } \nonumber \\
&&  \div(\overline{\rho} \bu) = 0 \mbox{   ,   } \nonumber  \\
&& \overline{\rho} \overline{T} \bu\cdot \grad \overline{s}  
= \div( \overline{k} \grad \tilde{T} ) \mbox{   ,   }
\nonumber  \\
&& \frac{\tilde{p}}{\overline{p}} = \frac{\tilde{\rho}}{\overline{\rho}} +  \frac{\tilde{T}}{\overline{T}}\mbox{   ,   }\nonumber \\
&& \grad^2 \tilde{\Phi} = 4\pi G \tilde{\rho} \mbox{   ,   }\nonumber \\
&& \curl(\bu \times \bB) = \curl( \overline{\eta} \curl \bB) \mbox{   ,   }
\nonumber  \\
&& \div \bB = 0\mbox{   ,   }
\label{eq:globallin}
\end{eqnarray}
where we approximated the equation of state with that of a perfect gas, 
and neglected perturbations to the chemical composition. Note that the 
latter approximation was chosen for the sake of simplicity: Wood \& McIntyre 
(2007) showed that the effects of compositional gradients (notably Helium) 
could play an important role in the tachocline dynamics. The 
background magnetic diffusivity $\overline{\eta}$, kinematic viscosity 
$\overline{\nu}$ and thermal conductivity $\overline{k}$ are calculated
from model S using the expressions provided by Gough (2007) (see Appendix A).

Note that the nonlinearities in the quantities relating to the 
magnetohydrodynamics of the interior (flow velocities and magnetic field) 
have not yet been tampered with\footnote{Also note that in principle, the thermal 
energy equation should contain an additional term, namely 
$\tilde{k} \grad \overline{T}$. This term is included in the numerical 
algorithm for the sake of completeness, but in practise has little influence 
on the result (at least in the case of the solar radiative zone). 
It will not be discussed in this paper for the sake of simplicity.}.
The next step is to use a spherical coordinate system ($r,\theta,\phi$) and 
move to a rotating frame of reference, by considering 
that $ \bu = \overline{\bu} + \tilde{\bu}$, with $\overline{\bu} = 
r \sin\theta \overline{\Omega} \ephi$ and $\tilde{\bu} = (\tilde{u}_r,\tilde{u}_\theta,\tilde{u}_\phi$). 
The value of $\overline{\Omega}$ 
adopted is uniquely defined by requiring that the total angular momentum of 
the convection zone be null in the rotating frame (Gilman {\it et al.}, 
1989): if $\Omega_{\rm cz}(\theta,r)$ is assumed to be independent of radius, 
and given by equation (\ref{eq:ocz}), then 
\begin{equation}
\overline{\Omega} = \Omega_{\rm eq} \left( 1 - \frac{a_2}{5} - \frac{3a_4}{35} \right) \mbox{  . }
\end{equation}
The equation for the conservation of momentum then becomes
\begin{eqnarray}
&& \overline{\rho} \tilde{\bu} \cdot \grad \tilde{\bu} +  2\overline{\rho} 
\overline{\bO} \times \tilde{\bu} + \overline{\rho} \overline{\bO} 
\times \overline{\bO} \times \br  \nonumber \\ 
= && 
- \grad \tilde{p} - \overline{\rho} \grad \tilde{\Phi}  - \tilde{\rho} 
\grad \overline{\Phi}  + \bj \times \bB + \div \Pi \mbox{   ,   }
\label{eq:rotating}
\end{eqnarray}
while the assumption of axial symmetry implies that the equations for the conservations of mass, 
thermal energy and magnetic flux are merely changed by replacing $\bu$ with $\tilde{\bu}$.

It is well known that the baroclinic deformation of the background state 
caused by the centrifugal force drives meridional flows, albeit very slow 
ones since the Sun is a slow rotator (Eddington, 1925; Sweet, 1950). Since 
the global turnover timescale of this Eddington-Sweet circulation is 
calculated to be orders of magnitude longer than the age of the Sun, 
it cannot play any role in the angular momentum or chemical mixing processes. 
In a steady-state calculation, however, these flows play an artificially 
important role and it is important to suppress 
them. This can be done by using the following momentum conservation 
equation instead (SZ92):
\begin{equation}
\overline{\rho} \tilde{\bu} \cdot \grad \tilde{\bu} + 
2\overline{\rho} \overline{\bO} \times \tilde{\bu}   =
- \grad \tilde{p} - \tilde{\rho} 
\grad \overline{\Phi}  + \bj \times \bB + \div \Pi \mbox{   ,   }
\label{eq:rotating2}
\end{equation}
where Cowling's approximation was also 
used to justify neglecting the perturbation 
of the gravitational potential related to density and temperature perturbations
caused by the remaining Coriolis and Lorentz forces.

Finally, we assume that the meridional flow velocities remain 
small (SZ92; GM98): we neglect quadratic terms in $\tilde{u}_r$ 
and $\tilde{u}_\theta$, but retain nonlinearities in 
$\tilde{u}_{\phi}$ to allow for significant differential rotation. 

To summarise, the model equations we consider are 
\begin{eqnarray}
&& \overline{\rho} \tilde{\bu} \cdot \grad \tilde{\bu} + 2 \overline{\rho} \overline{\bO} \times \tilde{\bu}   =
- \grad \tilde{p} - \tilde{\rho} \grad \overline{\Phi}  
+ \bj \times \bB + \div \Pi \mbox{   ,   }\nonumber  \\
&&  \div(\overline{\rho} \tilde{\bu}) = 0 \mbox{   ,   } \nonumber  \\
&& \overline{\rho} \overline{T} \tilde{\bu}\cdot \grad \overline{s}  
= \div( \overline{k} \grad \tilde{T})\mbox{   ,   } \nonumber  \\
&& \frac{\tilde{p}}{\overline{p}} = \frac{\tilde{\rho}}{\overline{\rho}} +  \frac{\tilde{T}}{\overline{T}}\mbox{   ,   }\nonumber \\
&& \curl(\tilde{\bu} \times \bB) = \curl( \overline{\eta} \curl \bB) \mbox{   ,   }
\nonumber  \\
&& \div \bB = 0\mbox{   ,   }
\label{eq:global2}
\end{eqnarray}
where the quantities solved for are the three components of $\tilde{\bu}$ (as defined above), the three 
components of $\bB = (B_r,B_\theta,B_\phi)$, and the three thermodynamical variables $\tilde{p}$, 
$\tilde{\rho}$ and $\tilde{T}$, and where quadratic terms in $\tilde{u}_r^2$, 
$\tilde{u}_\theta^2$ and $\tilde{u}_r\tilde{u}_\theta$ are neglected (see Appendix B for the full set of 
equations expanded in a spherical coordinate system). 

This set of equations is equivalent to the one used by GM98 before 
they apply their boundary-layer analysis. It also 
reduces to the one used by SZ92 in the non-magnetic case,
provided the following further approximations are applied: the Boussinesq 
approximation, assuming that $\tilde{\Omega} \ll \overline{\Omega}$ and 
that $\tilde{u}_\theta \gg \tilde{u}_r$. Finally, it is a steady-state and
axisymmetric version of the equations used by BZ06.

For comparison, note that R\"udiger \& Kitchatinov (1997), 
MacGregor \& Charbonneau (1999), and Kitchatinov \& R\"udiger (2006) neglect 
meridional flows entirely in their calculations of the structure of the 
radiative interior\footnote{although in work of Kitchatinov \& R\"udiger (2006), the effect 
of the flows does influence the boundary conditions applied to the magnetic field.} Sule, 
Arlt \& R\"udiger (2005) include 
meridional flows but assume that the poloidal field is fixed, so that the
interaction between the field and the flows through the Lorentz force in their 
paper is fundamentally linear (and therefore inappropriate).  

\subsection{Domain boundaries and boundary conditions}
\label{subsec:numbcs}

\subsubsection{Computational domain}

\begin{figure}
\epsfig{file=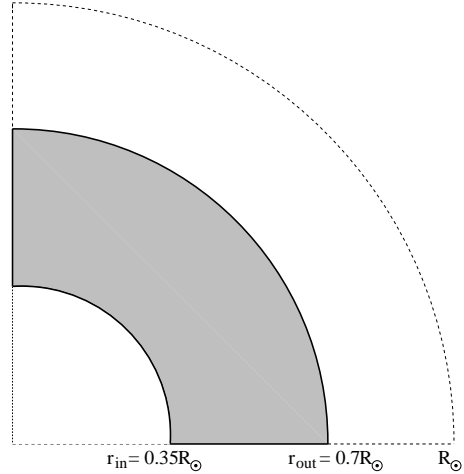,width=6cm}
\caption{The computational domain is limited to the spherical shell
shown  here in grey.}
\label{fig:modelfig}
\end{figure}

The geometry of the
computational domain is shown in Figure \ref{fig:modelfig}. 
The upper boundary is selected to be slightly below the base of 
the convection zone, at $r_{\rm out} = 0.7\rsun$. 
The lower boundary is at $r_{\rm in} =
0.35\rsun$ as in BZ06. The inner core (for $r < r_{\rm in}$) 
can be viewed as a solid metallic sphere, within which a
permanent magnetic dipole is maintained. The selection of the position of the
lower boundary was found to have little influence on the tachocline dynamics 
for low enough values of the diffusivities.

\subsubsection{Model boundary conditions}
\label{subsec:modelbc}

The selection of adequate boundary conditions for this model 
is a very delicate task. Given the elliptic and nonlinear nature of the system 
considered, the existence of solutions is not guaranteed, and solutions
(when they exist) are entirely controlled by the boundary conditions applied. 

Our aim is to propose a set of boundary conditions at the upper
boundary of the computational domain which reproduces most faithfully the
influence of the convection zone on the radiative zone, while those 
near the lower boundary are selected in such a way as to impose the 
required dipolar field, but otherwise be as inconspicuous as possible.

The set of coupled partial differential equations described in 
(\ref{eq:global2}) and expanded in Appendix B represents a 
12th order system, and therefore requires 12 independent boundary conditions.
\\ 
\\
{\it Bottom boundary conditions.} The bottom boundary of the computational domain is located at 
$r_{\rm in}$, and should be thought of as the interface between the modelled 
fluid and an electrically and thermally conducting solid sphere 
(hereafter ``the core''). Impermeability implies
\begin{equation}
\tilde{u}_r = 0  \mbox{   at   } r = r_{\rm in} \mbox{   .   }
\end{equation}
We generally impose no-slip boundary conditions at $r_{\rm in}$ (except when 
specifically mentioned), so that 
$\tilde{u}_\theta = 0$ and $\tilde{u}_\phi = r_{\rm in} \sin \theta \tilde{\Omega}_{\rm in}$, 
where $\tilde{\Omega}_{\rm in}$ is the constant angular velocity of the core in the rotating 
frame. The value of $\tilde{\Omega}_{\rm in}$ is selected in such a way as to guarantee that 
the total torque applied to the core be zero:
\begin{equation}
\int_{-\pi/2}^{\pi/2} \left( \overline{\rho} \overline{\nu} r^2_{\rm in} \sin^2 \theta \frac{\partial \tilde{\Omega}}{\partial r} + r_{\rm in} \sin\theta \frac{B_r B_\phi}{4\pi} \right) \sin\theta \dd \theta = 0 \mbox{   .  }
\end{equation} 

Assuming that the core is a homogeneous thermally 
conducting solid implies that the temperature fluctuations within 
satisfy Laplace's equation
\begin{equation}
\grad^2 \tilde{T}^{\rm in} = 0 \mbox{   .  }
\end{equation}
The regular solution can be expanded upon standard 
eigenfunctions, namely
\begin{equation}
\tilde{T}^{\rm in}(r,\theta) = \sum_{n = 0}^{\infty} a^{\rm in}_n P_n\left(\cos \theta\right) r^n \mbox{   ,}
\label{eq:Tin}
\end{equation}
where $P_n$ is the $n$-th order Legendre polynomial, and the set $\{a_n^{\rm in}\}$ are integration constants.
Requiring the simultaneous continuity of $\tilde{T}$ and its derivative 
at $r = r_{\rm in}$ while satisfying equation (\ref{eq:Tin}) constrains the 
 $\{a_n^{\rm in}\}$ and yields a unique relationship between $\tilde{T}$ and 
$\partial \tilde{T}/ \partial r$, which is used as a boundary condition 
on temperature.  

Two boundary conditions for the magnetic field are required. They are 
obtained in a similar fashion to that of the temperature fluctuations, by 
assuming that the core
is a homogeneous electrically conducting solid. In that case 
the magnetic field within satisfies
\begin{equation}
\grad^2 \bB^{\rm in} = 0 \mbox{   ,   }
\end{equation}
which, by axial symmetry, is equivalent to 
\begin{equation} 
(\grad^2 \bB^{\rm in})_\phi = 0 \mbox{   and   } (\curl \bB^{\rm in})_\phi = 0 \mbox{   .  }
\end{equation}
Defining a flux function $\chi$ as 
\begin{equation}
\bB^{\rm in}_{\rm p} = \curl\left( \chi^{\rm in} \ephi \right) 
\end{equation}
implies that both $B^{\rm in}_\phi$ and $\chi^{\rm in}$ satisfy the same equation, 
namely
\begin{equation}
\grad^2 B^{\rm in}_\phi - \frac{B^{\rm in}_\phi}{r^2 \sin^2\theta} = 0 \mbox{  , similarly for   } \chi^{\rm in}\mbox{   .   }
\end{equation}
The regular solution for $B^{\rm in}_\phi$ can be expanded as
\begin{equation}
B_\phi^{\rm in}(r,\theta) = \sum_{n=1}^\infty b^{\rm in}_n \frac{\partial}{\partial \theta} \left( P_n(\cos\theta)\right) r^n\mbox{   ,   }
\label{eq:Bphiin}
\end{equation} 
where as before the $\{b_n^{\rm in}\}$ are integration constants.
Requiring the simultaneous continuity of $B_\phi$ and its derivative 
at $r = r_{\rm in}$ while satisfying equation (\ref{eq:Bphiin}) 
yields a unique relationship between $B_\phi$ and 
$\partial B_\phi/ \partial r$, which is used as one of the two required 
boundary conditions on the magnetic field.

A point-dipole located at $r=0$ is required 
to maintain the primordial magnetic field in this steady-state study. Note that
this implicitly assumes that the dynamics of the tachocline occur
on timescales much shorter than the global magnetic diffusion time 
(which is a reasonable assumption since the global magnetic diffusion 
timescale is of the order of the age of the Sun). To match 
$\bB^{\rm in}_{\rm p}$ to a point dipole of 
amplitude $B_0$ at $r=r_{\rm in}$ and $\theta =0$ (on the polar axis), we 
require that
\begin{equation}
\chi^{\rm in} = B_0 \frac{r^3_{\rm in}}{2} \frac{\st}{r^2} + \sum_{n=1}^\infty c^{\rm in}_n \frac{\partial}{\partial \theta} \left( P_n(\cos\theta)\right) r^n\mbox{   .   }
\label{eq:chiin}
\end{equation}
Requiring the simultaneous continuity of $B_r$ and $B_\theta$  
at $r = r_{\rm in}$ provides a relationship between these
two quantities, which is used as the second of the two required
boundary conditions on the magnetic field.  \\ 
\\ 
{\it Top boundary conditions.} The choice of boundary conditions near 
the top boundary should model  the effects of the convective zone dynamics 
on the radiative--convective interface.

Requiring the continuity of the angular velocity profile across the interface 
is a natural choice for the boundary condition on $\tilde{u}_{\phi}$, 
although one must bear in mind that it is intrinsically equivalent to 
assuming that the interface is a no-slip boundary. For consistency, 
a similar condition should then also be imposed on the other 
component of the meridional flow parallel to the boundary, $\tilde{u}_\theta$. 
Hence we select
\begin{eqnarray}
\tilde{u}_\theta(r_{\rm out}, \theta)  &=& u^{\rm out}_\theta(\theta) \mbox{   ,   }\nonumber \\
\tilde{u}_\phi(r_{\rm out}, \theta) &=& r_{\rm out}\st \left(\Omega_{\rm cz}(\theta)-\overline{\Omega}\right)\mbox{   ,   }
\end{eqnarray}
where $u^{\rm out}_\theta(\theta)$ could be any desired latitudinal velocity
profile, and $\Omega_{\rm cz}$ is given by equation (\ref{eq:ocz}).

Boundary conditions are also needed on the vertical velocity $\tilde{u}_r$. 
In previous models of the radiative zone (Paper I; BZ06), the 
radiative--convective interface was assumed to be impermeable. However, 
this boundary condition is inadequate as a model of a (permeable) 
fluid interface (see Section \ref{subsec:sourceprob}). Instead, we 
consider the more general distribution: 
\begin{equation}
\tilde{u}_r (r_{\rm out}, \theta)  = u^{\rm out}_r(\theta) \mbox{   ,   }
\end{equation}
where the profile selected for $u^{\rm out}_r(\theta)$ is discussed in 
more detail in Section \ref{subsubsec:urforcing}.

The temperature boundary condition near the upper boundary is selected 
to model a very efficiently conducting fluid, by requiring that 
\begin{equation}
\nabla^2 \tilde{T}^{\rm out} = 0
\end{equation}
outward of the computational domain. Matching $\tilde{T}$ to $\tilde{T}^{\rm out}$ 
provides a unique relationship between $\tilde{T}$ and $\partial \tilde{T}/\partial r$ 
at the outer boundary. 

The boundary conditions for the magnetic field are obtained by assuming that 
the convection zone is an infinitely diffusive medium (so that $\nabla^2 \bB^{\rm out} = 0$ 
outward of $r_{\rm out}$) and selecting the solution which decays as $r \rightarrow \infty$: 
\begin{eqnarray}
&& B_\phi^{\rm out}(r,\theta) = \sum_{n=1}^\infty b^{\rm out}_n \frac{\partial}{\partial \theta} \left( P_n(\cos\theta)\right) r^{-(n+1)} \mbox{   ,   }\nonumber \\
&& \chi^{\rm out}(r,\theta) = \sum_{n=1}^\infty c^{\rm out}_n \frac{\partial}{\partial \theta} \left( P_n(\cos\theta)\right) r^{-(n+1)} \mbox{   .   }
\end{eqnarray}
Matching these solutions at $r = r_{\rm out}$ with the solution for $\bB$ within the domain 
provides relationships between $B_\phi$ and its derivative, and between $B_r$ and $B_\theta$ 
respectively.

\subsection{Non-dimensional parameters, and the nature of boundary layers}
\label{subsec:bls}

\subsubsection{Enhanced diffusivities}
\label{subsubsec:diffs}

The typical solutions of the set of equations and boundary conditions 
described in Sections \ref{subsec:numeqs} and \ref{subsec:numbcs} 
are known to contain thin boundary layers and internal layers of 
various kinds, typically scaling as some (positive) 
power of the diffusivities. 
The limited achievable numerical resolution of the algorithm used implies 
that the set of equations (\ref{eq:global2}) 
cannot be solved for solar values of these quantities. To address the problem
we multiply the diffusivities $\overline{\nu}(r)$, $\overline{\eta}(r)$ and 
\begin{equation}
\overline{\kappa}(r) = \frac{\overline{k}}{\overline{\rho} \overline{c}_{\rm p}} 
\end{equation}
by the enhancement factors $f_\nu$, $f_\eta$ 
and $f_{\kappa}$ respectively. When these factors are all selected equal 
to each other the Prandtl and magnetic Prandlt numbers
(Pr = $\overline{\nu}/\overline{\kappa}$ and Pr$_m = \overline{\nu}
/\overline{\eta}$ respectively) 
are solar everywhere. Note that we specifically do not view the numerical model
diffusivities as ``turbulent diffusivities''. We strive instead to derive
characteristic scalings of the solutions with each of the $f-$factors, and 
understand the asymptotic behaviour of the solutions as they are simultaneously 
reduced towards unity. 

\subsubsection{Non-dimensional parameters}
\label{subsubsec:params}

A better understanding of the scaling behaviour of the numerical solutions 
as $f_\nu$, $f_\eta$ and $f_\kappa$ 
are varied can be achieved by working with 
non-dimensional parameters. In what follows, we scale all distances with 
the solar radius $\rsun$, and all velocities with $\rsun\Omega_{\rm eq}$ 
(see Appendix C for details). The magnetic field strength is scaled by $B_0$, 
which is the amplitude of the imposed field on the inner boundary on the 
polar axis 
(see equation (\ref{eq:chiin})). Within this framework, the following
non-dimensional numbers naturally emerge: 
\begin{eqnarray}
&& E_{\nu} = \frac{f_\nu \overline{\nu}}{\rsun^2 \Omega_{\rm eq}} = f_\nu E_\nu^{\odot} \mbox{   ,   } \nonumber \\
&& E_{\eta} = \frac{f_\eta \overline{\eta}}{\rsun^2 \Omega_{\rm eq}} = f_\eta E_\eta^{\odot} \mbox{   ,   } \nonumber \\
&& E_{\kappa} = \frac{f_\kappa \overline{\kappa}}{\rsun^2 \Omega_{\rm eq}} = f_\kappa E_\kappa^{\odot} \mbox{   ,   }\nonumber \\
&& \Pr = \frac{E_\nu}{E_\kappa} = \frac{f_\nu}{f_\kappa} \Pr^{\odot} \mbox{   ,   }\nonumber \\
&& {\rm H} = \sqrt{\frac{E_\eta E_\nu}{\Lambda}} = \sqrt{f_\nu f_\eta} {\rm H}^\odot \mbox{  where   }  \Lambda  = \frac{B^2}{ 4\pi \rho_0 \rsun^2 \Omega_{\rm eq}^2} \mbox{   ,   }\nonumber \\  
&& {\rm Bu} = \left(\frac{ND_\rho}{\rsun \Omega_{\rm eq}}\right)^2 = {\rm Bu}^{\odot} \mbox{   ,   }
\end{eqnarray} 
where $D_\rho$ is the local density scaleheight ($D_\rho = 8.6 \times 10^9$ cm in the tachocline region) and $N$ is the buoyancy frequency ($N = 8 \times 10^{-4}$ s$^{-1}$ in the tachocline region). All of the numbers defined above 
actually vary with position within 
the solar interior. The first three are Ekman numbers, 
defined as the ratios of the rotation timescale to the respective 
diffusion timescales. 
The solar Ekman numbers calculated using the microscopic diffusivities 
at the base of the convection zone are $E_\nu^\odot \simeq 2 \times 10^{-15}$, 
$E_\eta^\odot = 3 \times 10^{-14}$, and
$E_\kappa^\odot = 10^{-9}$ respectively, all well below unity. The fourth is the 
Prandtl number, equal to $2\times 10^{-6}$ near the base of the convection zone in the Sun. 
The Hartmann number H depends naturally on the magnetic field strength $B$ selected. 
For $B = 1$G, ${\rm H}^\odot = 5 \times 10^{-9}$ at the base of the convection zone, 
and H scales as the inverse of the magnetic field strength. The Burger number Bu 
is of the order of $10^3$.

\subsubsection{Expected boundary layers and desirable parameter range}
\label{subsubsec:paramsreg}

To run simulations in a parameter regime that is at least in the same
asymptotic limit as the Sun, one should bear in mind the following 
constraints. To be in the non-diffusive regime, one must ensure that all three
Ekman numbers in the simulations are well-below unity. The 
$f-$parameters selected for the numerical calculations should therefore
remain below $(E_\nu^{\odot})^{-1}$, $(E_\eta^{\odot})^{-1}$ and 
$(E_\kappa^{\odot})^{-1}$ respectively, and also ensure that the hierarchy
$E_\nu \ll E_\eta \ll E_\kappa$ is respected. 

In the non-magnetic case, GB08 showed that the
characteristic lengthscales of the dynamics of the radiative zone 
(for solar parameter values) can be summarised as $l_1 \sim \rsun$ 
(i.e. the entire interior), $l_2 \sim \frac{D_\rho}{\sqrt{\Pr {\rm Bu}}}$ 
(i.e. $l_2 \sim 3 \sqrt{\frac{f_\kappa}{f_\nu}} \rsun$) and $l_3 \sim 
E_\nu^{1/2} \rsun$. Hence, to retain the same ordering of lengthscales as 
in the Sun ($l_2 > l_1 \gg l_3$), the factor $\sqrt{f_\kappa/f_\nu}$ must 
not drop below 1/3, placing 
very strong constraints on the Prandtl number used in the simulations 
(i.e. Pr should not exceed $\sim$10Pr$^{\odot}$).

The typical magnetic boundary layers are reviewed by Acheson \& Hide (1973). 
In the case of a magnetic field 
parallel to the boundary considered, the nature of the boundary layer depends on 
the respective sizes of $E_\nu$ and H: if $E_\nu >$ H then 
the boundary layer thickness is of the order of 
H$^{1/2} \rsun$, while if $E_\nu  < H$ then the boundary layer thickness scales as 
$E_\nu^{1/2} \rsun$ (i.e. the boundary layer is actually of Ekman type).
We conclude that for realistic field strengths, the boundary layers 
in the horizontal field case (e.g. near the equator) will always be of 
Ekman type. When the magnetic field is oblique, the situation changes: 
the Ekman regime is recovered when $E_\nu \ll {\rm H}^2$, while the boundary 
layer has a magnetic nature when $E_\nu \gg {\rm H}^2$, in which case its thickness
is of the order of H$\rsun$. We therefore conclude that unless the magnetic field 
strength (in the tachocline) has amplitudes lower than a mG, one
may actually expect a Hartmann layer near the radiative--convective interface.

To guarantee that $E_\nu/ {\rm H}^2 \gg 1$ in the simulations we
need $f_\eta \ll E_\nu^\odot / ({\rm H}^\odot)^2 \sim 10^2$ if 
the magnetic field is indeed 
of the order of 1G as suggested by GM98. This is clearly
not an achievable goal, so we must inflate the magnetic field strength 
artificially to be of the order of 1T near the outer boundary;  in that case 
the constraint on $f_\eta$ is much less stringent, with
$f_\eta \le 10^{10}$.

Finally, if we rely on flows downwelling from the convection zone to confine the 
magnetic field, then the amplitude of the flows assumed at the radiative--convective 
interface must be large enough for their magnetic Reynolds number ${\rm R}_m$ to be 
larger than unity. Since ${\rm R}_m$ is very loosely defined as 
\begin{equation}
{\rm R}_m = \frac{|u_r| \Delta}{f_\eta \overline{\eta}}\mbox{   ,   }
\label{eq:Rm}
\end{equation}
where $|u_r|$ is the typical amplitude of the radial flows and $\Delta \sim 0.03 \rsun$ 
is the observed thickness of the tachocline (both in cgs units), then equation (\ref{eq:Rm}) 
implies that we need a priori
flows with amplitude of the order of 
$f_\eta \overline{\eta} / \Delta \sim 2000 (f_\eta / 10^{10}) $cm/s 
to have a magnetic Reynolds number of order one. For values of $f_\eta \sim 10^{10}$ 
(typically achieved in the simulations), this naturally requires unphysically large 
flow amplitudes. The consequences of this choice are discussed in Sections 
\ref{subsec:explor} and \ref{sec:disconcl}.

\subsection{Numerical method}
\label{subsec:numalgo}

\subsubsection{Method and tests.}

The numerical method selected for the solution of this elliptic system
is based on the expansion of the governing equations and boundary conditions 
upon a truncated set of orthogonal functions spanning the $\theta-$direction 
(Chebishev polynomials to be precise), followed by the solution of the 
associated set of ordinary differential equations (ODEs) in the $r$-direction 
using a second-order 
Newton-Raphson-Kantorovich (NRK) relaxation 
algorithm. The numerical algorithm 
is globally similar to the one used by Garaud (2001) and 
in Paper I, and is summarised in Appendix C. The high latitudinal resolution 
required to capture the detailed 
structure of the interior dynamics demands a high-order expansion in $\theta$. 
This in turn requires a sophisticated parallelized version of NRK
An outline of RELAX -- the parallel NRK algorithm freely available 
upon request to P. Garaud -- 
is given in Appendix D (see also Garaud \& Garaud, 2007). 

The complexity of the full set of equations (\ref{eq:global2}) 
precludes the derivation of analytical solutions in the general case, 
and hence direct tests of the full numerical algorithm. However,
we have extensively and successfully tested the code 
through direct comparisons of the numerical solutions with 
analytical solutions of selected {\it subsets} of the system: Paper I 
tested the un-stratified magnetic version of the code, while
GB08 tested the stratified non-magnetic version of the 
code.

\subsubsection{Note on the convergence of the solutions}
\label{subsubsec:convsol}

In the NRK/RELAX algorithm, a guess is required for the solution of the
ODE system, and refined by successive iterations until the desired accuracy 
is reached. The convergence of the algorithm is guaranteed and immediate 
for linear systems regardless of the initial guess, but requires 
more accurate initial guesses for increasingly dominant 
nonlinearities. As a result, 
solutions are found in practise, first for high-values of the 
diffusivities (i.e. high $f$) and with a low number of latitudinal modes,
and then progressively followed into the nonlinear regime by lowering $f$
and increasing the number of modes.  

Convergence of the solutions usually becomes painfully difficult as $f$ 
is lowered below a certain threshold. The causes of the problem can be varied: 
in some cases, the nonlinearities begin to 
dominate the solution and bifurcations occur which reduce the basin of 
attraction of the solution followed; in others, the spatial structure of the 
solution becomes finer and finer, requiring more latitudinal modes and 
more radial meshpoints to be fully resolved.  

The only way to address the first problem is to follow the solution 
carefully, and reduce $f$ only by a very small fraction between each run. 
This has the disadvantage of being a frustratingly slow process, with 
little to be gained in practise. As a matter of fact, 
the likely change of stability of the solution tracked (suggested by the 
change in convergence properties) is an interesting dynamical information 
in itself (see Section \ref{subsubsec:algoselec}).  

To address the second problem typically requires increasing the number of modes
and meshpoints, which leads to a very rapid increase in computational costs.  
Nonetheless, some acceptable trade-offs are found with typical simulations
 using a few thousand meshpoints (mostly concentrated in the boundary layers) 
and about 60-100 latitudinal modes. These require no more than a few hours 
per iteration on about $2^7$-$2^8$ processors (see Appendix D).

\subsubsection{Discussion of the selection of the algorithm}
\label{subsubsec:algoselec}

At this point it may be worth discussing an oft-raised question about the 
selection of this specific numerical method of solution, as an 
alternative to the time-dependent ASH code 
used by BZ06. Both methods, when taken individually, 
are equally useful tools for studying the problem, each with their
own advantages and limitations. 

For example, the algorithm of BZ06 studies the general properties 
of the radiative zone under the forcing considered, and in addition 
provides information on the stability of the system, 
naturally taking into account 
transport by smaller-scale three-dimensional flows when 
necessary. However, 
the very lengthy integration time required 
implies that only very few different parameter 
values can realistically be explored. Moreover, 
the time-dependent nature of the algorithm makes it more difficult 
to identify quasi-steady states (in particular 
when the typical timescales in the numerical model are far less 
obviously separated than in the Sun).

By contrast, the simplified axisymmetric and steady-state nature of the 
equations solved in Paper I and here implies that each solution 
can be found in at most a day of real-time computations (more typically 
an hour). As a result, we are able to explore a 
wide range of parameters and study in detail the behaviour of the solutions
as a function of the imposed forcing (boundary flows, 
differential rotation, magnetic field strength, ...) and as a function 
of the background model (diffusivities, thermo-dynamical profile, ...). 
This feature is invaluable when trying to understand how solutions scale 
with the various parameters -- see GB08 for example.
In addition, we always know that the solutions found
are indeed the steady-state solutions of the system, and not transients. 
On the other hand, taking into account transport
by three-dimensional turbulence is not possible (without artificial
parameterization of its effects). In addition, we cannot 
{\it a priori} know whether the steady-states found are stable, or 
whether they may be unstable to either 2D or 3D perturbations.

Interestingly, however, we can make use of the fact that 
the emergence of unstable solutions implies a reduction of the 
basin of attraction of the stable ones (see Section \ref{subsubsec:convsol}), 
and (tentatively) 
identify sudden convergence difficulties 
with the presence of a bifurcation. For example, 
the simulations of BZ06 can be compared with ours for 
\begin{equation}
f_\nu = 2.6\times 10^8 \mbox{  , } f_\eta = 8 \times 10^7 \mbox{   and } f_\kappa = 8 \times 10^5\mbox{   ,   }
\label{eq:bzf}
\end{equation}
(see Section \ref{subsec:bz}). Our numerical algorithm has little difficulty 
converging to a solution 
for parameters $f_\nu$, $f_\eta$ and $f_\kappa$ respectively about
2 times larger than the ones quoted above. To go from there 
down to the values used by BZ06 on the other hand requires much more care,
a problem which is very likely related to the fact that the 
BZ06 solution is seen to be mildly unstable to 2D and 3D perturbations.

\section{Numerical experiments}
\label{sec:numexp}

The versatility of the numerical algorithm constructed combined with its rapid
real-time execution (see Appendix D) enables us to perform an extensive 
range of simulations, both in terms of the parameter regime studied and in 
terms of the forcing applied. In this section we 
use this tool in a number numerical experiments.

In Section \ref{subsec:bz}, we first prove our diagnostic for the 
failure of previous numerical attempts (Paper I, BZ06) to reach 
a dynamical equilibrium qualitatively similar to the one studied by GM98.
We reproduce a quasi-steady state similar to the one achieved 
by BZ06, and study its scaling properties. We show that the 
meridional flow velocities induced in the tachocline, when the
radiative--convective interface is modelled as an impermeable boundary, 
are always too low to confine the field.

We contrast this result in Section \ref{subsec:pump} with a similar 
calculation in which meridional flows are directly pumped through the 
boundary (through some unspecified radial forcing mechanism). We show that 
their amplitude remains sufficiently high in the tachocline to provide 
significant field confinement.

We then perform a suit of simulations in Section
\ref{subsec:explor},  in the ``desirable'' parameter regime specifically 
selected in Section \ref{subsubsec:paramsreg}, 
and discuss the numerically derived solutions both qualitatively and 
quantitatively. We show that the system can, in some situations and for 
low-enough diffusivities, achieve a balance qualitatively similar to that of 
the GM98 model, but that the solutions also reveal a number of previously 
unaccounted-for dynamical subtelties (see Sections \ref{subsec:fieldconf}, 
\ref{subsec:omvsf} and \ref{subsec:ekmanrole}).

\subsection{Failure of ``impermeable'' simulations}
\label{subsec:bz}

\subsubsection{The calculation}

In this first numerical experiment, we compare our quasi-steady solutions 
with those integrated by BZ06. 
Note that since the overall interior magnetic field 
strength decays slightly 
in their time-dependent calculations, while our 
steady-state solutions fix the value of the field amplitude near 
the inner boundary, the two simulations can only be compared qualitatively 
at best. Nonetheless, the comparison is meaningful 
since the system is thought to relax to a quasi-steady 
equilibrium on a timescale comparable with the meridional 
circulation timescale, which is argued (by BZ06) to be 
much shorter than the magnetic diffusion timescale.

To achieve the same parameter regime as the one selected 
by BZ06 in terms of the diffusivities, we set 
\begin{equation}
f_\nu = 2.6\times 10^8  \tilde{f} \mbox{  , } f_\eta = 8 \times 10^7  \tilde{f} \mbox{   and } f_\kappa = 8 \times 10^5  \tilde{f}
\label{eq:bzf2}
\end{equation}
and progressively reduce $\tilde{f}$ from about $10^5$ to 1. When $\tilde{f} = 1$, 
our diffusivities are the same in the tachocline region as 
theirs\footnote{Also note that the BZ06 simulations actually have constant 
diffusivities throughout the radiative interior, while ours vary with radius
as in the Sun. Since we only seek a qualitative comparison of the 
two simulations, this difference in the diffusivities near the inner 
boundary is irrelevant.}. Naturally, we select $u_r^{\rm out} = 0$ to mimic
an impermeable boundary as in BZ06. We also modify our bottom boundary 
conditions to be ``stress-free'', again to be consistent with their simulations. 
Finally, the radial component of the magnetic field on the inner boundary 
is set to be same as the initial condition selected by BZ06 (i.e. $B_r = 500$G 
at the poles at $r = r_{\rm in}$).

\begin{figure}
\epsfig{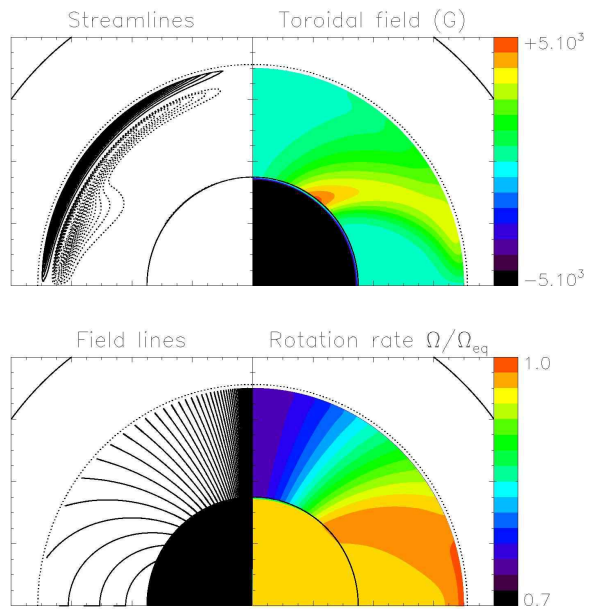}
\caption{Quasi-steady state results for a numerical simulation similar to that of 
Brun \& Zahn (see Section \ref{subsec:bz}). In each quadrant, the dotted circle 
marks the position of the base of the convection zone at $r = r_{\rm cz}$, the 
outer solid circle marks the solar surface and the inner solid circle is at 
$r = r_{\rm in}$. {\it Top left:} Streamlines of the meridional flow. Clockwise 
flows are shown as solid lines, and anticlockwise flows are shown as dotted lines. 
{\it Top right:} Toroidal field (in Gauss). {\it Bottom left:} Magnetic field lines. 
{\it Bottom right:} Angular velocity profile. Note how the magnetic field lines 
connect to the bottom of the convection zone and establish a differentially 
rotating Ferraro state throughout the domain.}
\label{fig:bzsimu}
\end{figure}

\subsubsection{Qualitative comparison of the results}

The results for $\tilde{f} = 1$  (see equation
(\ref{eq:bzf2})) are shown in Figure \ref{fig:bzsimu}.  The overall
structure of the streamlines and of the magnetic fields  lines,
as well as the amplitude of the toroidal field and of the differential
rotation compare qualitatively well with
the simulations of BZ06 (see the $t=4.7$Gyr solution in Figures 4 and
5 of BZ06). Small discrepancies in  the magnetic field geometry can be
attributed to the fact that their magnetic  field  is diffusing out of
their inner core, while it is maintained in our simulation by a
permanent dipole.

This steady-state calculation therefore confirms the main conclusion of the 
study carried out by BZ06, namely that in the case where the radiative--convective 
interface is modelled as an impermeable boundary, the only possible quasi-steady 
solution has an unconfined field structure leading to a 
differentially rotating Ferraro state.

\subsubsection{The nature and amplitude of the flows}

\begin{figure}
\epsfig{file=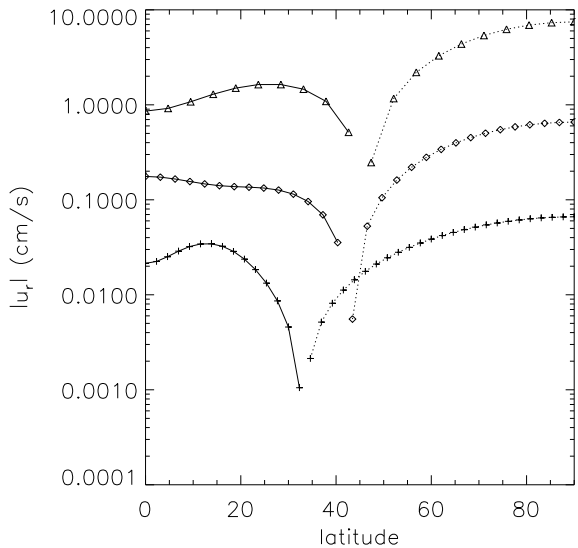,width=8cm}
\caption{Absolute value of the radial component 
of the meridional flow  at $r = 0.68 \rsun$ as a function of latitude, when 
$\tilde{f} =1$ (plus symbols), $\tilde{f} =  10$ (diamonds), and $\tilde{f} = 100$ 
(triangles), for $\tilde{f}$ defined in equation (\ref{eq:bzf2}). 
The radial position 
was selected to be safely below the Ekman layer for all three calculations. The 
solid part of the curve denotes $u_r > 0$ and the dotted part of the curve denotes 
$u_r < 0$. Note how $|u_r|$ roughly scales with $\tilde{f}$.}
\label{fig:bzur}
\end{figure}

We now justify our diagnostic of the failure of ``impermeable simulations'' 
to find a confined-field solution. The induced meridional flows in 
this set of simulations are generated primarily by Ekman pumping on 
the boundary. Indeed, the magnetic field strength selected by BZ06 implies 
a field of the order of a few tens of Gauss at the most near the outer 
boundary, and therefore according to the analysis of 
Section \ref{subsubsec:paramsreg} a boundary layer in the Ekman regime rather 
than in the Hartmann regime. In that case, the flow amplitudes are in fact
well-estimated by the (non-magnetic) work of GB08 and 
scale as $E_\nu / \sqrt{\Pr}$. This is verified in Figure \ref{fig:bzur}, 
which 
shows the radial velocity profile of the flows as a function of latitude, well below 
the Ekman layer (at $r = 0.68\rsun$), 
in three different simulations: for $\tilde{f} = 1$, $\tilde{f} = 10$ and 
$\tilde{f} = 100$. It is clear from the figure that
the meridional flow velocity decreases linearly with $\tilde{f}$, 
and can in fact be shown to scale as predicted above. 

As a result, one can straightforwardly see that the magnetic Reynolds number 
does not increase with decreasing $E_\eta$ when the respective diffusivity 
ratios are held constant: with R$_m$ defined as in equation (\ref{eq:Rm}),  
if $|u_r|$ is proportional to $\sqrt{f_\kappa f_\nu}$ then 
the magnetic Reynolds number is in fact proportional to 
$\sqrt{f_\nu f_\kappa}/f_\eta$ and remains roughly constant as 
$\tilde{f}$ decreases. The effective magnetic Reynolds number 
of the BZ06 simulations can be estimated from the flow amplitudes calculated 
in  Figure \ref{fig:bzur} to be ${\rm R}_m \sim 5 \times 10^{-3}$ -- 
far too low to influence the magnetic field.

Using the true solar diffusivities would not change the situation 
since in that case it can be shown that the expected magnetic Reynolds number 
is of the order of ${\rm R}_m \sim 4 \times 10^{-2}$. 
We have therefore confirmed 
our original diagnostic, namely that the reason for the failure of 
numerical simulations to find confined-field solutions is not related to 
the fact that 
the diffusivities in the simulations are too high (i.e. selecting solar 
diffusivities would yield a similar result), but rather to
a fundamental mis-representation of the radiative--convective interface. 

\subsection{The possibility of magnetic field confinement by the thermo-viscous mode}
\label{subsec:pump}

Following the original idea proposed by GM98, we now explore 
the possibility of confining the interior field through flows which 
are directly downwelling from the convection zone. As shown by GB08, 
flows which are pumped into and out of the radiative 
zone (by stresses in the convection zone for example) can retain significant 
amplitudes throughout much of the tachocline (and below), and could therefore
result in a magnetic Reynolds number larger than unity for low enough 
diffusivities. We illustrate this idea with a qualitative example, which,  
for ease of comparison with the results of the previous section,
is selected to have the same diffusivity ratios and magnetic field 
strength as those of BZ06. 

\subsubsection{Structure of the imposed meridional flow}
\label{subsubsec:urforcing}

We now select a radial velocity profile $u^{\rm out}_r(\theta)$ which
satisfies the following properties: (i) $u^{\rm out}_r(0) = 0$  (the
radial velocity is  zero on the polar axis), and the flow is symmetric with
respect to the equator (ii) the total mass flux through the boundary (at $r =
r_{\rm cz}$) is zero:
\begin{equation}
\int_{0}^{\pi/2} u^{\rm out}_r(\theta) \sin\theta {\rm d}\theta = 0\mbox{   ,   }
\end{equation}
 and (iii) the total angular momentum flux carried by the meridional circulation  
through the boundary is zero:
\begin{equation}
\int_0^{\pi/2} r^2 \sin^2\theta \tilde{\Omega}_{\rm cz}(\theta) u^{\rm out}_r(\theta) {\rm d} \theta  = 0\mbox{   .   }
\end{equation}
The simplest solution which
 satisfies all three constraints is a 6-th order polynomial in  $\mu = \cos(\theta)$:
\begin{equation}
u^{\rm out}_r(\mu) = U_0 \left( 1 + \sum_{n=1}^3 \alpha_n \mu^{2n}\right) 
\label{eq:urout}
\end{equation}
 where 
\begin{eqnarray}
\alpha_1 = - \frac{3 ( 715 + 169 a_2 + 141 a_4)}{ 143 + 13 a_2 + 25 a_4} \mbox{   ,   }\nonumber \\
\alpha_2 = \frac{5 ( 1001 + 299 a_2 + 207 a_4)}{ 143 + 13 a_2 + 25 a_4}\mbox{   ,   } \nonumber \\
\alpha_3 = - \frac{91 ( 33 + 11 a_2 + 7 a_4)}{ 143 + 13 a_2 + 25 a_4} \mbox{   ,   }
\end{eqnarray}
and $U_0$ is the radial velocity at the equator.

\begin{figure}
\epsfig{file=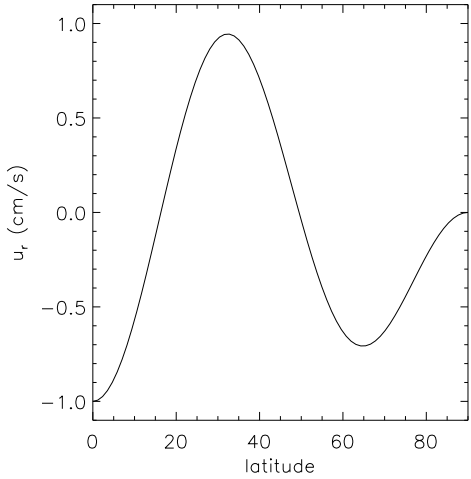,width=7cm}
\caption{Example of imposed radial flow $u^{\rm out}_r$ as a function of 
latitude, for the values of $a_2$ and $a_4$ described in equation 
(\ref{eq:ocz}), satisfying the conditions listed in the main text. 
The velocity of the flow ($U_0$) 
here is arbitrarily selected to be $-1$ cm/s at the equator. 
The width of the upwelling region, as can be seen in the 
figure, spans about 30$^\circ$ in latitude and is centred on 30$^\circ$.}
\label{fig:uforcing}
\end{figure}

An example of this profile can be seen in Figure \ref{fig:uforcing}: for the 
values of $a_2$ and $a_4$ given in (\ref{eq:ocz}) the upwelling region is 
centered at $30^{\circ}$ latitude, and ranges from $15^{\circ}$ to $45^{\circ}$. 
Note that the selection of the ``simplest'' profile is fairly arbitrary -- had 
we chosen a different set of orthogonal functions, we would have obtained a 
slightly different flow profile. One may also wish instead to select a profile 
where the width and position of the upwelling region is prescribed, or instead 
of selecting the polar velocity to be zero, to select the equatorial velocity 
to be zero, etc... The only 
necessary constraints, however, are zero mass and angular momentum fluxes.
 These conditions are often ignored, but are crucial to obtaining
meaningful results; failure to satisfy them yields unphysical 
angular velocity profiles (see Garaud, 2007).

\subsubsection{Qualitative structure of the solutions}

\begin{figure}
\epsfig{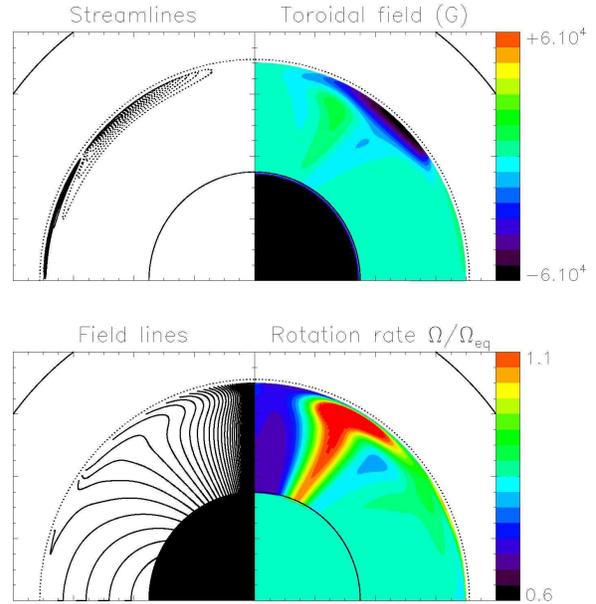}
\caption{Same as \ref{fig:bzsimu}, but with an imposed meridional flow at 
$r = r_{\rm out}$ (see Figure \ref{fig:uforcing}).}
\label{fig:bzwithu}
\end{figure}

In the simulation presented here, the diffusivities are selected as in 
equation (\ref{eq:bzf2}) with $\tilde{f} = 1$. 
A radial meridional flow profile is imposed at the outer boundary (see 
equation (\ref{eq:urout})), with an amplitude $U_0$ which is progressively 
increased from 0 to about -5.5 cm/s. 
The resulting numerical solution is presented in Figure 
\ref{fig:bzwithu}. It clearly shows that the field lines are 
strongly distorted by the meridional flows, and seem to be confined both 
near the equator, and more importantly also near the poles. In  fact, our results
appear to be at least qualitatively similar to the analytical 
solutions of Wood \& McIntyre (2007), which focussed on the polar regions only.
The angular velocity profile is very different from the one observed in the 
simulations shown in Figure \ref{fig:bzsimu} and reveals uniform rotation 
from the equator up to about 60$^{\circ}$ in latitude. The rotation rate of 
the inner core is found to be 0.86$\Omega_{\rm eq}$. 

We have therefore shown that, at least on a qualitative basis, magnetic field
confinement is possible provided a radial flow is imposed near the radiative--convective interface. 

\subsection{Exploration of parameter space}
\label{subsec:explor}

Having proved on a qualitative basis that the GM98 model 
may indeed lead to global field confinement, we
now turn to a more quantitative analysis of the resulting tachocline
dynamics. For this purpose, we return to using the boundary conditions
discussed in Section \ref{subsec:numbcs}.

In what will be referred to from here on as the ``fiducial model'',  
the enhancements factors
$f_\nu$, $f_\eta$ and $f_\kappa$ are now selected as follows to be in the 
``desirable'' parameter range discussed in Section \ref{subsubsec:paramsreg}: 
\begin{equation}
f_\nu = \frac{\tilde{f}}{10} \mbox{   ,   } f_\eta =  \tilde{f} \mbox{
,  } f_\kappa= \frac{\tilde{f}}{100}\mbox{   ,   }
\end{equation}
and the factor $\tilde{f}$ is progressively reduced from $10^{15}$ down
to $10^{10}$ typically ($8\times 10^9$ for the lowest case). For
$\tilde{f} =  10^{10}$, and in the tachocline region, $E_\nu  \sim  2
\times 10^{-6}$, $E_\eta \sim 3 \times 10^{-4}$ and $E_k \sim 10^{-1}$. In
that case, and for all simulations, $\Pr = 10 \Pr^{\odot}$ (guaranteeing
that the thermo-viscous mode lengthscale is indeed of the order of
the entire radiative zone), and $\Pr_m = 0.1 \Pr_m^{\odot}$. Note that the 
simulations of BZ06 use a Prandtl number of the order of 300$\Pr^{\odot}$, 
which does not satisfy our criterion for the hierarchy of the expected 
characteristic lengthscales. 

The magnetic field strength on the inner boundary at the pole is selected
to be $B_0 = 7$T, so that the field strength in the tachocline would be 
of the order of 1T in the absence of meridional flows. The radial
flow velocity at the outer boundary is the one discussed in Section 
\ref{subsubsec:urforcing} with $U_0 = -550$cm/s, and the latitudinal 
flow velocity is everywhere zero.

\begin{figure}
\epsfig{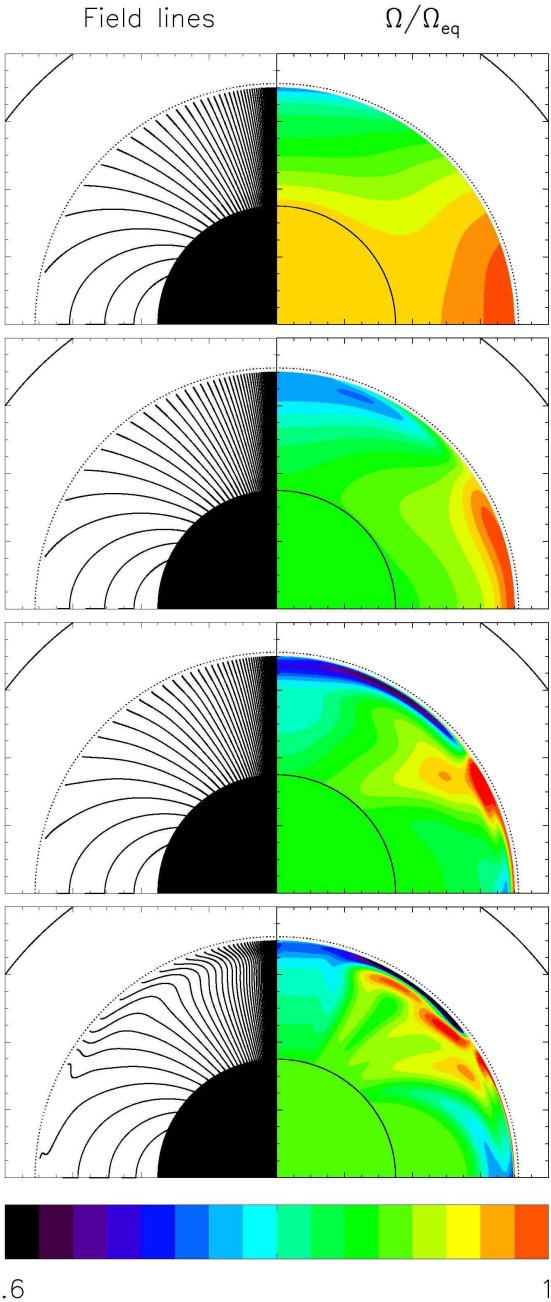}
\caption{Evolution of the poloidal field topology and of the angular 
velocity profile in the fiducial model for, from top to bottom, 
$\tilde{f} = 10^{13}$, $10^{12}$, $10^{11}$ and $10^{10}$. Note 
that regions with $\Omega < 0.6 \Omega_{\rm eq}$ are drawn in black.}
\label{fig:fvaryombp}
\end{figure}
\begin{figure}
\epsfig{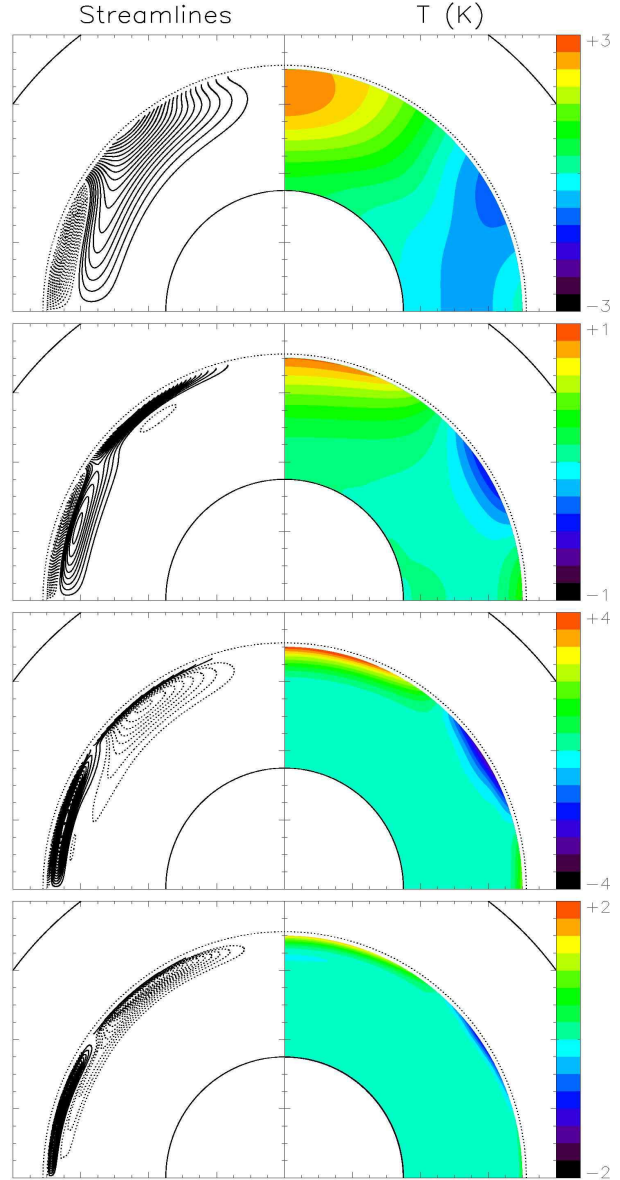}
\caption{Evolution of the topology of the streamlines and of the 
temperature perturbations in the fiducial model for, from top to 
bottom, $\tilde{f} = 10^{13}$, $10^{12}$, $10^{11}$ and $10^{10}$. 
Note that the streamlines shown are representative streamlines, 
the contours selected are different in each plot. Also note that 
the temperature scale is different in each plot.}
\label{fig:fvarypsit}
\end{figure}

\subsubsection{From the diffusive regime toward the asymptotic regime}

The qualitative evolution of the poloidal field topology and of 
the angular velocity profile as $\tilde{f}$ is reduced 
can be seen in Figure \ref{fig:fvaryombp}, while the corresponding figure 
for the streamlines and the temperature profile is Figure \ref{fig:fvarypsit}. 
In all simulations, the viscous Ekman number is much smaller than one. 
The magnetic field begins to affect the angular velocity profile as
$\tilde{f}$ decreases from $10^{13}$ to $10^{12}$, notably in the deeper 
interior\footnote{Recall that the magnetic diffusivity is smaller 
in the deeper interior, see Appendix A.}. This transition occurs when 
the magnetic Ekman number drops below one. The structure of the 
temperature perturbation profile changes as $\tilde{f}$ is decreased 
from $10^{12}$ down to $10^{11}$, which is attributed this time to 
the ``thermal'' Ekman number approaching one. Finally, there is a clear 
transition  between the ``unconfined field'' configuration 
for $\tilde{f}=10^{11}$ and the ``confined field'' configuration for 
$\tilde{f} = 10^{10}$, which corresponds to the magnetic Reynolds number 
increasing to values above unity. 

\begin{figure}
\epsfig{file=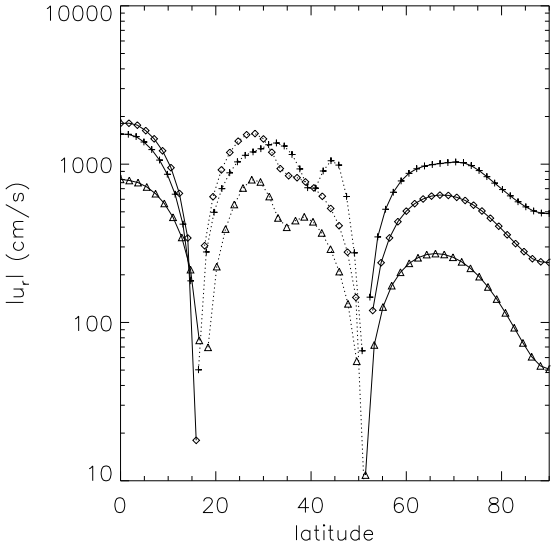,width=8cm}
\caption{Radial flow velocities at $r=0.68\rsun$ as a function of 
latitude in the fiducial model, for three values of $\tilde{f}$: 
$\tilde{f} = 10^{10}$ (plus symbol), $\tilde{f} = 3\times 10^{10}$ 
(diamonds) and $\tilde{f} = 10^{11}$ (triangles). As in Figure 
\ref{fig:bzur}, the solid lines denote upward flow ($\tilde{u}_r > 0$) 
and the dotted lines denote downward flow ($ \tilde{u}_r < 0$). The 
figure therefore shows a downwelling in mid-latitude, with upwelling 
at the poles and the equator. Note how the amplitude of the radial 
velocity appears to increase with decreasing $\tilde{f}$.}
\label{fig:urfid}
\end{figure}

Contrary to the impermeable-boundary case studied by BZ06, the radial 
flow velocities do not decrease with the diffusivities, but are actually 
seen to increase as $\tilde{f}$ is reduced. This is illustrated in Figure 
\ref{fig:urfid}, which shows the radial flow velocity near the base of 
the presumed tachocline (at $r=0.68 \rsun$, as in Figure \ref{fig:bzur}), 
as a function of latitude, for three values of $\tilde{f}$. Provided 
$E_\kappa < 1$, it appears that $\tilde{u}_r \propto 1/\tilde{f}$ 
although this scaling cannot, for numerical reasons, be explored 
over many orders of magnitude. As a result, the nonlinear interaction 
of the flow and the field rapidly becomes stronger and the field appears 
to be more and more confined, as can be seen in the last panel of Figure 
\ref{fig:fvaryombp}. 

\subsubsection{The ``lowest diffusivities'' simulation}

\begin{figure}
\epsfig{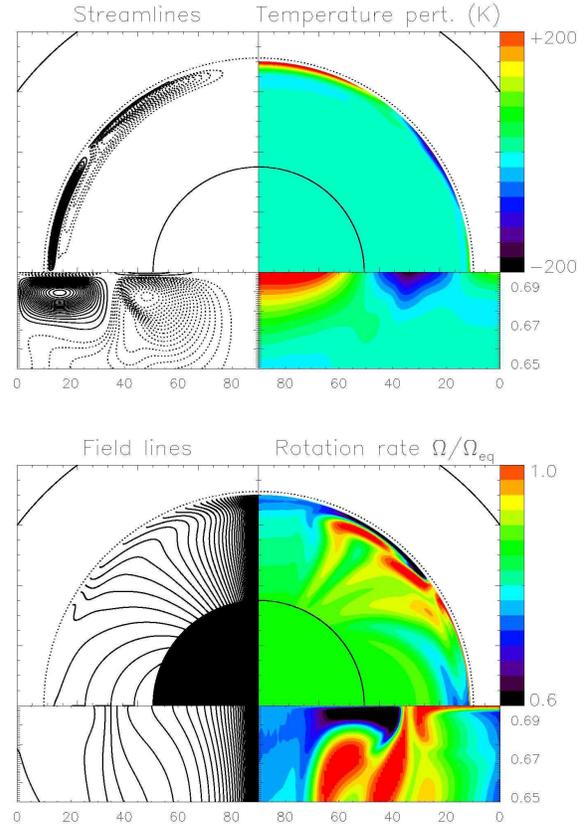}
\caption{Results of the simulations of the fiducial model for 
$\tilde{f} = 8\times 10^9$. For each quantity plotted (from top 
left to bottom right, respectively, representative streamlines, 
temperature perturbations, field lines and rotation rate), we show 
both a full quadrant as well as a zoomed-in strip of outer boundary 
layer (the region between $r = 0.65 \rsun$ and $r=r_{\rm out}$). As 
usual, the solid streamlines denote clock-wise flows, and dotted 
streamlines denote counter-clockwise flows. Note that regions with 
$\Omega < 0.6 \Omega_{\rm eq}$ are drawn in black.}
\label{fig:bestsim}
\end{figure}

We now decrease $\tilde{f}$ as far as possible, until convergence becomes
too difficult. The lowest value achieved in the fiducial model is 
$\tilde{f} = 8\times 10^9$ and the 
results are presented in Figure \ref{fig:bestsim}. The 
number of radial meshpoints used is 3000, and the number of latitudinal modes 
is 80 (note that since only even modes are selected to guarantee equatorial 
symmetry, solving for 80 modes means that the highest Fourier mode considered 
actually is $\cos(158 \theta)$, see Appendix C). 

The top-right quadrant of Figure \ref{fig:bestsim} 
shows representative streamlines. The Ekman layer near the outer boundary, 
of width $\sim 0.001 \rsun$, is just visible in the top-right strip. 
The Ekman layer flow does indeed downwell at the poles and equator, 
and upwells in mid-latitudes. 

More clearly visible is the ``thermo-viscous'' mode below the 
Ekman layer, which 
(rather unexpectedly, to be honest) has the opposite 
structure: downwelling in mid-latitudes and upwelling near the poles and 
equator. This mode appears to be confined roughly 
within $r \in [0.67,0.7] \rsun$, an
effect which can only be attributed to the Lorentz forces arising from 
the magnetic field underneath. Indeed, the solutions of GB08
in the absence of magnetic fields in the same parameter regime do not
reveal any structure on this typical lengthscale. The associated 
effect of the flow in confining the magnetic field is also obvious 
in the bottom-left quadrant: the field lines are most 
strongly distorted away from the strictly dipolar structure in the same 
region ($r \in [0.67,0.7] \rsun$). Field confinement is discussed in more 
detail in Section \ref{subsec:fieldconf}. 

The solution for the angular velocity profile is again more complex than 
expected: the 
combination of the two modes of propagation of the meridional flows creates
radial structures both on the Ekman scale and on a larger scale 
($r \sim 0.1 \rsun$), as well as latitudinal structures which are clearly 
seen to be correlated with the magnetic field lines. However, contrary to the 
solution shown in Figure \ref{fig:bzsimu}, there is no clear evidence for a 
{\it large-scale} latitudinal or radial gradient in $\Omega$ below 
about 0.6$\rsun$, including in the polar regions. This is the first set of 
self-consistent simulations to reveal this kind of solution. Since the 
angular velocity profile inferred from helioseismic inversions can be thought
of as a weighted spatial average of the true angular velocity profile 
in the Sun, and since the spatial extent of the averaging kernels is 
much larger than the features seen in the simulations, the angular 
velocity profile shown in Figure \ref{fig:bestsim} would 
be observationally interpreted to be uniform below $0.6\rsun$,
whereas the one shown in Figure 
\ref{fig:bzsimu} would not. The actual value of the rotation rate
calculated in the simulations is discussed in Section \ref{subsec:omvsf}.

A closer look at the angular velocity profile in the region 
$r \in [0.65,0.7] \rsun$ (see the bottom-right strip)
reveals a change in the dominant angular momentum transport processes between
the uppermost layers ($r > 0.67\rsun$) and the lower layers ($r < 0.67 \rsun$).
When $r > 0.67\rsun$ the angular velocity contours are more-or-less
aligned with the streamlines, whereas for $r < 0.67 \rsun$ the angular velocity
contours are aligned with the magnetic field lines. This is at least
qualitatively consistent with the idea proposed by GM98 of a multi-layered
tachocline, in which the dynamics of the uppermost layer are controlled by 
the tachocline flows, while the dynamics of the bulk of the radiative zone 
are controlled
by the primordial field. The interface between the two regions, which could
be identified with the ``magnetic diffusion layer'' of the GM98 model, clearly 
has quite a complex topology.

One can also observe a number of very strong localised 
``jets'' of more rapidly or more slowly rotating fluid. The jets appear to 
be stronger for lower diffusivities (one can easily compare the  
results of Figure \ref{fig:bestsim} with the last panel of Figure 
\ref{fig:fvaryombp}). In fact, we tentatively attribute the numerical scheme's 
failure to find solution for $\tilde{f} < 8 \times 10^9$ to an 
intrinsic instability of the jets. 
But whether such strong features would exist in the Sun is debatable: both 
the imposed flow velocities and the magnetic field strength have been 
artificially increased by many orders of 
magnitude to ensure a magnetic Reynolds number greater than one in the 
simulations, and a Hartmann number closer to that of the Sun (see Section 
\ref{subsubsec:paramsreg}). As a result, the amplitude of these jets is also 
artificially increased in the simulations compared with what one may expect 
in the Sun. Unfortunately, given the nonlinear nature of the problem, 
it is difficult 
to predict what the actual strength of these jets would be should 
diffusivities, magnetic field and imposed flows velocities be 
simultaneously reduced to the expected solar values. 

\subsection{Magnetic field confinement}
\label{subsec:fieldconf}

To study the nature of field confinement more quantitatively, we
consider the ratio $\xi = |B_r / B_{r,{\rm dipolar}}|$ where
$B_{r,{\rm dipolar}}$ is the hypothetical radial component of the
solution for the magnetic field in the purely diffusive case:
\begin{equation}
B_{r,{\rm dipolar}}(r,\theta) =  B_0 \cos{\theta}
\left(\frac{r}{r_{\rm in}}\right)^{-3} \mbox{  . }
\end{equation} 
Typically, we expect that $\xi < 1$ when magnetic field lines are
either completely expelled from a region, or bent to the
horizontal. On the other hand, we expect that $\xi > 1$ in regions
where the magnetic field lines are pushed together by converging
(latitudinal) flows. It is therefore a good diagnostic  of the
processes we are interested in.

\begin{figure}
\epsfig{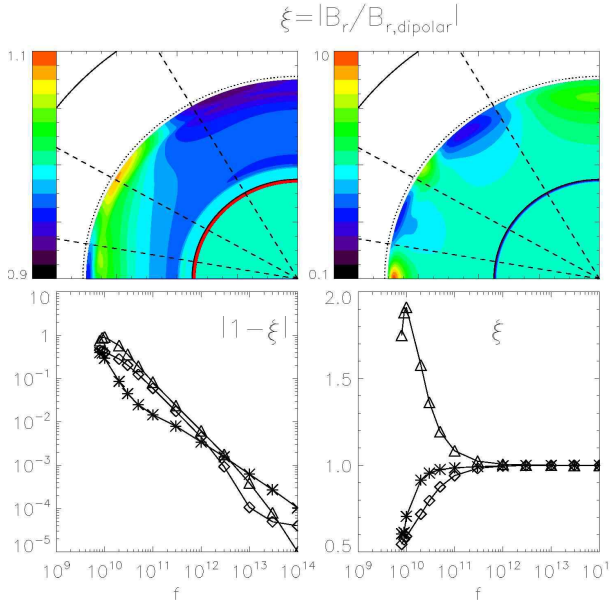}
\caption{Measure of magnetic confinement, using the quantity $\xi$ as
defined in the main text, Section \ref{subsec:fieldconf}. {\it Top
left}: Variation of $\xi$ with radius and latitude, in a fiducial
model for $\tilde{f} = 10^{11}$. {\it Top right}: Same as on the left
but for $\tilde{f}=8\times 10^9$. Note the change in the scale of the
perturbations between the two plots, and also note that here the
contours are logarithmically spaced. {\it Bottom left}: $|1-\xi|$
evaluated at the outer boundary, at three different latitudes
respectively in the downwelling regions ($10^{\circ}$, plus symbols,
and $60^{\circ}$, diamonds) and in the middle of the upwelling region
($30^{\circ}$, triangles). This log-log plot emphasises the power-law
relationship between $|1-\xi|$ and $\tilde{f}$ for ${\rm R}_m < 1$,
see main text. {\it Bottom right:} Similar to the left-side plot but
showing $\xi$ with a linear scale, emphasising the breakdown of this
relationship when ${\rm R}_m > 1$.}
\label{fig:fielddiff}
\end{figure}

The quantity $\xi$ is shown in Figure \ref{fig:fielddiff} for various
simulations. The top two quadrants are derived from the numerical
solutions for $\tilde{f} = 10^{11}$ and $\tilde{f} = 8\times 10^9$
respectively, and reveal a strong change in the behaviour of the
solutions when ${\rm R}_m < 1$ and ${\rm R}_m > 1$.

In the case where $\tilde{f} = 10^{11}$, we observe a good match
between the spatial variation of $\xi$ and the imposed forcing: $\xi <
1$ in downwelling regions and $\xi > 1$ in the upwelling regions (at
mid-latitudes). However, the amplitude of $\xi$ remains close to one,
revealing only a very weak effect of the flow on the field. This is
not entirely surprising since the magnetic Reynolds number for
$\tilde{f} = 10^{11}$ can be deduced from the flow amplitude shown in
Figure \ref{fig:urfid} to be ${\rm R}_m \sim 0.03$ at
$r=0.68\rsun$. In fact, in this very weakly nonlinear regime there is
some evidence for a correlation (see Figure \ref{fig:fielddiff},
bottom left quadrant) between $1-\xi$ and $\tilde{f}$: we find that
$|1-\xi| \sim 10^{10} \tilde{f}^{-1}$.

However, Figure \ref{fig:fielddiff} also reveals that matters become
more complex as ${\rm R}_m$ increases above unity. For $\tilde{f} =
8\times 10^9$ the magnetic Reynolds number is now of the order of
${\rm R}_m \simeq 1.5$, and the connection between the field and the
meridional flow is correspondingly fully nonlinear. The correlation
between $1-\xi$ and $\tilde{f}$ breaks down, as seen in the bottom
right quadrant. This incidentally proves that the kind of boundary
conditions advocated by Kitchatinov \& R\"udiger (2006) to mimic
``field confinement by meridional flows'' cannot accurately describe
the nonlinear dynamics of the system. We also observe that the regions
of enhancement and reduction of the radial field are no longer
perfectly correlated with the {\it imposed} upwelling/downwelling
regions. In fact, some enhancement in the radial field strength can be
seen near the equator as well as near the polar regions. This can be
attributed to the convergence points of the meridional flows {\it
below} the Ekman layer, which, as seen in Figure \ref{fig:bestsim}
have the opposite sign as that of the imposed flow.

All in all, although there is definite evidence for magnetic field
confinement along the lines of the model proposed by GM98, it is
probably fair to say  that the system behaves in a far more complex
way than anticipated,  and that the simulations in the fiducial model
are just beginning to  scratch the surface of the ``interesting''
regime.

\subsection{Interior angular velocity}
\label{subsec:omvsf}

\begin{figure}
\epsfig{file=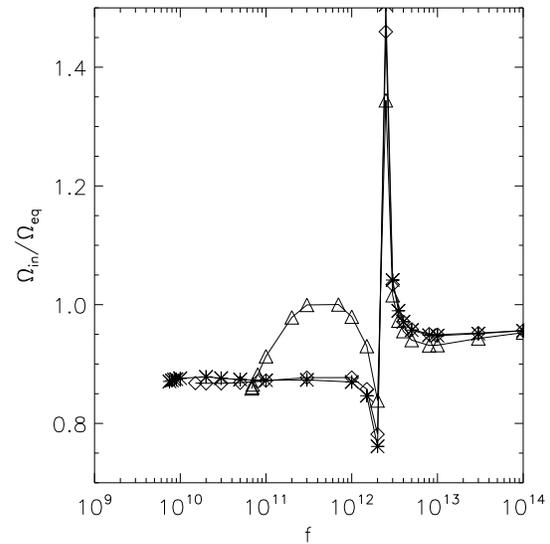,width=8cm}
\caption{Angular velocity of the inner core as a function of
$\tilde{f}$, in the fiducial model (star symbols), and in a model
where the amplitude of the imposed radial flow is ten times larger
(triangular symbols) and ten times smaller (diamond symbols)
respectively.  }
\label{fig:omvsf}
\end{figure}

As in Paper I, we consider the angular velocity of the inner core
$\Omega_{\rm in} = \overline{\Omega} + \tilde{\Omega}_{\rm in}$ (which
is an  eigenvalue of the calculation performed) as a diagnostic of the
most important angular momentum transport processes, and of the
``accuracy'' of the model  in reproducing the observations. This
quantity is shown in Figure  \ref{fig:omvsf} as a function of
$\tilde{f}$ for the fiducial model, and for simulations with faster 
and slower imposed flows.

In very diffusive simulations, the predicted angular velocity
$\Omega_{\rm in} \simeq 0.957 \Omega_{\rm eq}$  is consistent with
purely viscous angular  momentum transport throughout the entire
interior (see Gough, 1985).  As $\tilde{f}$ is progressively reduced,
the system undergoes a rather  impressive bifurcation at $\tilde{f}
\simeq 2.5 \times 10^{12}$, which  corresponds to the change of nature
of the boundary layer near the inner  core from an Ekman layer to an
Ekman-Hartmann layer. When this happens,  a new flow cell of opposite
vorticity appears right against the inner core,  so that angular
momentum transport changes direction. This leads to a rather dramatic
jump in the inner core rotation rate, which is unlikely to be of any
relevance to the Sun, but deserved an explanation.  Note that the
critical value of $\tilde{f}$ for which the bifurcation occurs  is
proportional to the imposed magnetic field strength $B_0$.

In the case of the fiducial model, the angular velocity of the core then 
remains roughly constant ($\Omega_{\rm in} \simeq (0.875 \pm 0.005) 
\Omega_{\rm eq}$) as 
$\tilde{f}$ is further reduced by two orders of magnitude. This range 
corresponds to the regime where ${\rm R}_m < 1$, where the magnetic 
field lines 
essentially remain dipolar. In fact, the same inner core velocity is also 
found for lower imposed flow velocities (see Figure \ref{fig:omvsf}), as well 
as for higher and lower field 
strength (always with ${\rm R}_m < 1$). This suggests that the value  
$ \Omega_{\rm in} \sim 0.875 \Omega_{\rm eq}$ is a rather universal 
characteristic of angular momentum transport by dipolar fields 
in the Hartmann regime, in 
this particular geometry, and subject to this particular imposed angular 
velocity profile at the outer boundary. 

However, it is clear from the analysis of Section \ref{subsec:fieldconf} 
that the simulations in the fiducial model have not quite yet 
reached the asymptotic 
regime even for the lowest values of $\tilde{f}$ achieved. As a matter 
of fact, in the last few points of the 
fiducial model curve (in the region where ${\rm R}_m$ begins to exceed one), 
$\Omega_{\rm in}$ is seen to change more rapidly, decreasing slightly 
below 0.87 $\Omega_{\rm eq}$.

The role of the meridional flows as angular momentum transporters can 
best be seen in a simulation with much higher imposed flow 
velocities (see the curve with the triangular symbols, for $U_0 = -5500$cm/s). 
In these simulations, the transition to ${\rm R}_m > 1$ 
occurs for values of $\tilde{f}$ typically 10 times larger 
than in the fiducial model. In that case, we observe that the predicted 
core velocity is 
notably different from the other two cases, and changes rather rapidly 
with $\tilde{f}$. It is therefore more than likely that the predicted 
angular velocity of the core in the fiducial model would also begin 
to deviate more noticeably away from $\Omega_{\rm in} = 0.875 \rsun$ were
we able to continue decreasing $\tilde{f}$. Again, we are only beginning to 
approach the asymptotic regime so that the simulated 
angular velocity of the core cannot yet and should not be compared 
directly with that of the Sun. 

\section{Discussion and prospects.}
\label{sec:disconcl}

\subsection{Summary of the results}

Despite the difficulties encountered when attempting to
find numerical solutions for asymptotically low values of the 
diffusivities, the nonlinear dynamics which emerge from our simulations 
are closer to what one may expect from the GM98 model than any other 
simulation performed to date. 

More precisely we do observe (see Section \ref{subsec:explor}) 
the quenching of the large-scale differential rotation by the 
primordial magnetic field, even in the polar regions.
We also observe the partial confinement of the field by the meridional flows
to the radiative interior (see Section \ref{subsec:fieldconf}), with 
the reduction of the radial field strength (in some regions)  
by more than 70\%. Finally, we observe the concurrent confinement of the 
meridional flows to the upper layers of the radiative zone 
($r > 0.67\rsun$) by the magnetic field. We have not yet been 
able to reduce the diffusivities down sufficiently far to observe
a truly segregated structure where the bulk of the tachocline flows
are completely magnetic-free. However, there is evidence in the 
observed rotation profile (see Figure \ref{fig:bestsim}) for a
transition between regions where angular-momentum transport is dominated
by the meridional flows, and regions where it is dominated by the magnetic 
field. The ``magnetic diffusion layer'' studied by GM98, 
which is thought to control this transition,
appears to have a rather complex geometrical structure which prevents a
more detailed study of the GM98 scalings.

The calculated value of the interior angular velocity $\Omega_{\rm in}$ 
(see Section \ref{subsec:omvsf}) does not match the observed value. 
However, since even the lowest-diffusivity simulations presented here are 
only just beginning to enter the 
asymptotic parameter regime described in Section \ref{subsubsec:paramsreg} 
we do not view the poor match with the observations as an intrinsic problem
with the GM98 model (yet) but rather as evidence that more work should be 
done to decrease the diffusivities even further -- a challenging task.

\subsection{The stability of the solutions?}

As described in Section \ref{subsec:explor}, the convergence of the solutions
becomes rather difficult when the values of the diffusivities are decreased 
below a certain threshold (in the fiducial model for 
$\tilde{f} < 8 \times 10^9$). The typical reasons for this convergence failure 
were listed and discussed in Section \ref{subsubsec:convsol}: 
intrinsic linear/nonlinear instabilities or insufficient latitudinal 
resolution. In this particular case, the various scenarios are
equally plausible, and difficult to disentangle without 
a supporting fully nonlinear 3D calculation. The field configuration 
near the inner core could be subject to instabilities (as seen in the 
simulations of BZ06). The Ekman(-Hartmann) layer near the outer boundary 
could also be subject to instabilities, or could be developing latitudinal 
structures which are too fine to resolve (see Figure \ref{fig:el}). 
We have in fact tentatively identified 
the emergence of strong jets in the angular velocity profile 
(see Figure \ref{fig:bestsim} for example) as the most 
likely source of instabilities (and convergence failure), but the reader 
should be cautionned that this statement is only speculative.

As mentioned in Section \ref{subsec:explor}, whether the equilibrium governed 
by the GM98 model is stable or unstable cannot directly be inferred from 
the stability of the numerical solutions (since they operate in a different 
parameter regime) but it is clearly a fundamental question. 
Should our diagnostic concerning the stability of the jets 
prove to be correct, then a possible path for 
future investigation (using this relaxation method) may be to continue
searching for solutions, starting where we left off, and slowly 
lowering simultaneously 
the amplitude of the imposed flow (and field) with the diffusivities to
limit the strength of the jets generated. This could be seen as blindly
navigating the stable regions of parameter space while avoiding instability 
reefs. But what could ideally be derived from such an exercise, 
should we be able to acquire sufficient data, are scaling relationships 
between the jet strength, the diffusivities and the imposed flow strength 
which may then be used to infer tentative information on the stability 
of the GM98 solution itself. 

\subsection{The role of an Ekman layer?}
\label{subsec:ekmanrole}

\begin{figure}
\epsfig{file=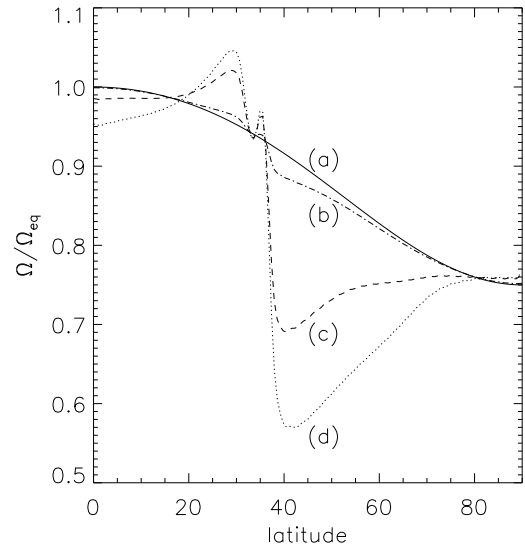,width=8cm}
\caption{Angular velocity profile on and just below the 
radiative--convective interface (a) at 0.7$\rsun$ (b) at 
0.6999$\rsun$, (c) at 0.6995$\rsun$ and (d) at 0.699$\rsun$ 
in the fiducial model for $\tilde{f}=8\times 10^9$. Note how 
the angular velocity profile in (and therefore also below) the 
Ekman layer is very different from the one imposed by the convection zone.}
\label{fig:el}
\end{figure}

Our simulations have also revealed a fundamental difference with 
the original GM98 model: the presence and role of an 
Ekman mode\footnote{Since the 
bulk of the GM98 tachocline is presumed to be magnetic-free, 
one should indeed consider an
Ekman mode rather than an Ekman-Hartmann mode}. While it is clear 
from Figure \ref{fig:bestsim} that the Ekman flows themselves play 
no role in confining the field (contrary to the claims made by 
Kitchatinov \& R\"udiger 2006) our simulations reveal that the role 
of the Ekman layer is still far from trivial.

Indeed, the differential 
rotation profile imposed at the radiative--convective interface 
is quite different from the differential rotation 
profile transmitted by the Ekman layer. This is illustrated
in Figure \ref{fig:el}, which shows the evolution of the angular 
velocity profile with depth across the Ekman layer. Thus while the GM98 model 
correctly describes
the non-viscous tachocline dynamics {\it below} the Ekman layer, 
the Ekman mode could in fact influence the system by 
modifying the angular velocity profile ``seen'' by the bulk of the 
tachocline\footnote{adding another layer to the sandwich...}. 

\begin{figure}
\epsfig{file=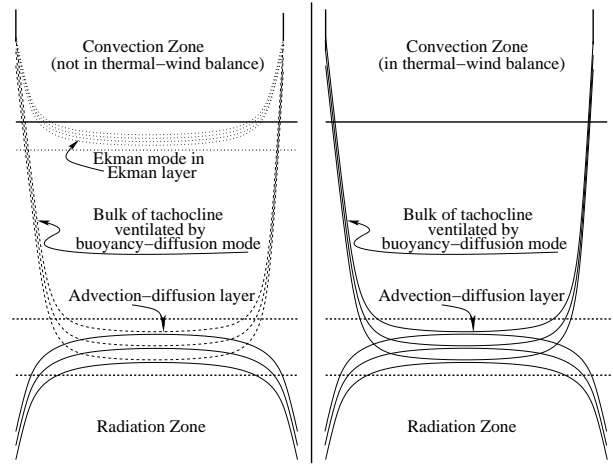,width=8cm}
\caption{\small A pictorial representation of the expected tachocline 
flows in two extreme cases. In both pictures, flows are downwelling 
from the convection zone into the radiative zone. As in the GM08 model, 
the bulk of the tachocline is ventilated by the thermo-viscous mode, 
which interacts with the magnetic field in a thin advection-diffusion 
layer. On the left, and as in the simulations presented in this paper, 
the convection zone is not dominated by thermal-wind balance, and a 
significant portion of the flows downwelling from the convection zone 
rapidly return within a thin Ekman layer. The angular velocity profile 
seen by the tachocline differs from the one observed in the convection 
zone. On the right, a hypothetical situation where the convection zone 
is essentially in thermal-wind balance. In that case, the Ekman mode 
is negligible and the GM98 model directly applies.}
\label{fig:bls}
\end{figure}

The extent to which this effect influences the tachocline is unclear. Firstly,
other processes which also transport angular momentum 
(convective overshoot, gravity waves, small-scale and large-scale magnetic 
stresses associated with the dynamo field) are known to take place in the close
vicinity of the radiative--convective interface. These processes were not
included in GM98's analysis and cannot be modelled with the 
present numerical algorithm, but could equally affect the tachocline dynamics.
Secondly, Ekman layers (laminar or turbulent) must {\it 
by definition} be present in any rotating system which exhibits very 
rapid changes in the imposed stresses. However, whether the Ekman 
mode actually plays an important role in the tachocline dynamics depends
on the relative importance of the bulk thermal-wind stresses and the 
combination of all the other rapidly varying turbulent stresses 
(see Figure \ref{fig:bls}). Indeed, if the base of the 
convection zone is already essentially in thermal-wind balance (see 
Miesch, 2005 for instance), then the amplitude of the Ekman mode and its effect
on the angular velocity profile will be small (GB08). On the other hand, 
if this is not the case then significant Ekman flows are expected 
(as in the simulations shown in the present work for example), with the aforementionned consequences.

\subsection{Prospects}

The present work has emphasized the necessity of flows downwelling 
from the solar convection zone as a means to confine the 
internal primordial field and guarantee the uniform rotation of the radiative
interior as originally proposed by GM98. The nature of the thermal 
and dynamical balance 
governing the convection zone itself therefore also controls 
the dynamics of the tachocline through the spatial variation and
amplitude of the flows crossing the radiative--convective interface. 
Meanwhile, steady progress in 
modelling the convection zone has emphasized its dependence on the thermal 
stratification of the tachocline (Rempel, 2005; Miesch, Brun \& Toomre, 2006). 
One can only conclude
that the dynamics of the convective and radiative regions are 
intrinsically and fundamentally coupled through this internal layer 
called the tachocline, and that the only sensible way forward from this 
point on is the construction of whole-Sun models including both regions.

\section*{Acknowledgements}

This work is the outcome of nearly ten years of discussions with many 
colleagues, including Nic Brummell, Sacha Brun, Gary Glatzmaier, Douglas Gough,
Michael McIntyre, Tamara Rogers, Nigel Weiss, Jean-Paul Zahn. 
P. Garaud thanks them all for their patience. Respective parts of this 
work have been 
supported, at various times, by New Hall (Cambridge), PPARC, and more recently
by Calspace (for J.-D. Garaud) and by NSF-AST-0607495 (for P. Garaud). The
numerical simulations were performed on the Pleiades cluster at UCSC, 
purchased using an NSF-MRI grant.

\appendix

\section[]{Selection of the background state}

\begin{figure}
\includegraphics[width=84mm]{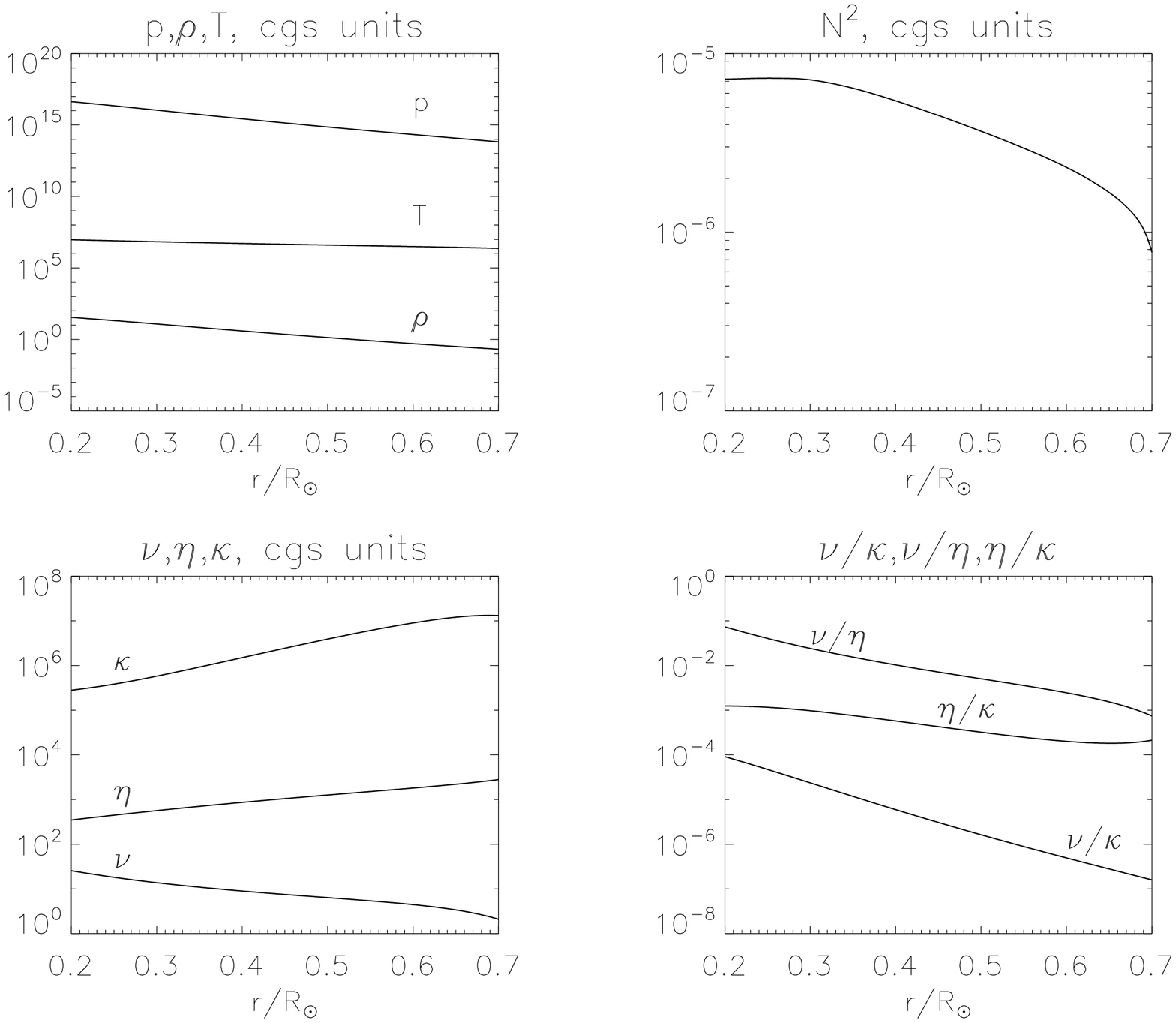}
\caption{Background pressure, density and temperature (top left corner), 
buoyancy frequency squared (top right corner), microscopic diffusivities 
(bottom left corner) and ratio of diffusivities (bottom right corner) 
from Model S of Christensen-Dalsgaard {\it et al.} (1991).}
\label{fig:background}
\end{figure}

The spherically symmetric background solar model selected for this
paper is Model S of Christensen-Dalsgaard {\it et al.} (1991). Model S  
provides calculated solar data for all of the relevant thermodynamical and
compositional quantities on a fixed mesh. In particular, we use the provided
fields $\overline{T}$, $\overline{p}$, $\overline{\rho}$, $\overline{g}$, 
$\overline{N}^2$ (where $N$ is the buoyancy frequency), $\overline{c}_{\rm p}$ 
and $\overline{\kappa}_{\rm R}$ (where $c_{\rm p}$ is the specific heat at 
constant pressure, and $\kappa_{\rm R}$ is the Rosseland mean opacity). 
In order to use this data for our purpose, we need to interpolate it
upon our own selected numerical mesh, using a standard rational
interpolation routine (cf. Numerical Recipes). The task is delicate
since any ``roughness'' in the interpolated functions results in failure
of convergence of the numerical algorithm, and the two numerical meshes are
intrinsically non-uniform: Model S has meshpoints strongly 
concentrated near 
the solar surface, near $r=0$ and at
the base of the convection zone $r_{\rm cz}  = 0.713 \rsun$, while
our numerical mesh has meshpoints strongly concentrated near the
domain boundaries $r_{\rm in}$ and $r_{\rm out}$. A new interpolating routine
was created which automatically selects a certain number of points from 
the original mesh, performs the rational interpolation upon the desired mesh, 
and checks for the smoothness of the interpolated function and of its derivative. 
Should the result be inadequate, the procedure is repeated with a different 
set of target points. Selected background quantities are shown in Figure 
\ref{fig:background}. 

The quantities $\overline{\nu}$, $\overline{\eta}$, and 
$\overline{k}$ are not provided by Model S, and must instead be calculated
using the formulae derived by Gough (2007):
\begin{eqnarray}
\overline{\nu}(r) &=& \nu_{\rm cz} \left[ 0.1 \left(\frac{\overline{T}}{T_{\rm cz}}\right)^4 \left(\frac{\overline{\rho}}{\rho_{\rm cz}}\right)^{-2} \left(\frac{\overline{\kappa}_{\rm R}}{\kappa_{\rm R,cz}}\right)^{-1} \right. \mbox{   ,   }\nonumber \\
 &+& \left. 0.9  \left(\frac{\overline{T}}{T_{\rm cz}}\right)^{5/2} \left(\frac{\overline{\rho}}{\rho_{\rm cz}}\right)^{-1} \left(\frac{\ln \overline{\Lambda}}{\ln \Lambda_{\rm cz}}\right)^{-1} \right] \mbox{   ,   }\nonumber \\
\overline{\eta}(r) &=& \eta_{\rm cz} \left(\frac{\overline{T}}{T_{\rm cz}}\right)^{-3/2} \left(\frac{\ln\overline{\Lambda}}{\ln \Lambda_{\rm cz}}\right) \mbox{   ,   }\nonumber \\
\overline{k}(r) &=& \overline{\rho} \overline{c}_{\rm p} \overline{\kappa}(r) \mbox{   ,   }
\end{eqnarray} 
where the Coulomb logarithm $\ln\overline{\Lambda}$ and the thermal 
diffusivity $\overline{\kappa}$ are given by 
\begin{eqnarray}
\overline{\Lambda}(r) &=& \Lambda_{\rm cz}\left(\frac{\overline{\rho}}{\rho_{\rm cz}}\right)^{-1/2} \left(\frac{\overline{T}}{T_{\rm cz}}\right)^{3/2} \mbox{   ,   }\nonumber \\
\overline{\kappa}(r) &=& \kappa_{\rm cz} \left(\frac{\overline{T}}{T_{\rm cz}}\right)^{3}\left(\frac{\overline{\rho}}{\rho_{\rm cz}}\right)^{-2} \left(\frac{\overline{\kappa_{\rm R}}}{\kappa_{\rm R,cz}}\right)^{-1} \mbox{   .   }
\end{eqnarray}
The following quantities are those at the base of the convection zone: $T_{\rm cz} = 2.3 
\times 10^6$K, $\rho_{\rm cz} = 0.21$g/cm$^3$, $\kappa_{\rm R,cz} = 19$cm$^2$/g, $\ln\Lambda_{\rm cz} = 2.5$, $\nu_{\rm cz} = 27$cm$^2$/s, $\eta_{\rm cz} = 410$cm$^2$/s and $\kappa_{\rm cz} = 1.4 \times 10^7$cm$^2$/s.

\section[]{Model equations in spherical coordinates}

The model equations described in the system (\ref{eq:global2}) are now 
expanded in a spherical coordinate system $(r,\theta,\phi)$, with 
$\tilde{\bu} = (\tilde{u}_r,\tilde{u}_\theta,\tilde{u}_\phi)$, $\bB = (B_r,B_{\theta},B_{\phi})$ and $\bj = (j_r,j_\theta,j_\phi)$.
\noindent The $r-$component of the momentum equation:
\begin{eqnarray}
- \overline{\rho} \st \left( 2\overline{\Omega} +  \frac{\tilde{u}_\phi}{r\st}\right) \tilde{u}_{\phi} &=& -\frac{\partial \tilde{p}}{\partial r} - \tilde{\rho} \frac{\dd \overline{\Phi}}{\dd r} + j_\theta B_\phi - j_\phi B_\theta \nonumber \\ &+& f_\nu (\div\Pi)_r\mbox{   .   }
\end{eqnarray}
The $\theta-$component of the momentum equation:
\begin{eqnarray}
- \overline{\rho} \ct \left( 2 \overline{\Omega} +  \frac{\tilde{u}_\phi}{r\st} \right) \tilde{u}_{\phi} &=& j_\phi B_r- j_r B_\theta \nonumber \\ &+& f_\nu (\div\Pi)_\theta\mbox{   .   }
\end{eqnarray}
The $\phi-$component of the momentum equation:
\begin{eqnarray}
\overline{\rho} \left(2 \overline{\Omega} + \frac{\tilde{u}_\phi}{r\st}\right) ( \ct \tilde{u}_\theta + \st \tilde{u}_r )  & = & j_r B_\theta - j_\theta B_r \nonumber \\
 &+& f_\nu (\div\Pi)_\phi\mbox{   .   }
\end{eqnarray}
where the role of $f_\nu$ was discussed in section \ref{subsubsec:diffs}.
The divergence of the viscous stress tensor, for a stratified fluid, 
can be derived from Batchelor (1994 edition, pp. 147 and 601).
\\
The mass conservation equation:
\begin{equation}
\frac{1}{r^2} \frac{\partial}{\partial r} \left( r^2 \overline{\rho} \tilde{u}_r \right) + \frac{\overline{\rho}}{r \st}\frac{\partial }{\partial \theta} \left (\st \tilde{u}_{\theta} \right) = 0 \mbox{   .   }
\end{equation}
The thermal energy equation can be re-written as (see SZ92)
\begin{equation}
\frac{\overline{\rho} c_{\rm p} \overline{T} N^2 }{g} \tilde{u}_r = \frac{f_\kappa}{r^2} \frac{\partial }{\partial r} \left( r^2 \overline{k} \frac{\partial \tilde{T}}{\partial r} \right)  + \frac{f_\kappa \overline{k}}{r^2 \st } \frac{\partial }{\partial \theta} \left( \st \frac{\partial \tilde{T}}{\partial \theta} \right) \mbox{   ,   }
\end{equation} 
where $ g = \dd \Phi/\dd r$ and $N$ is the Brunt-Vaisala (buoyancy) frequency.

\noindent The equation for the conservation of magnetic flux can be integrated once, and axial symmetry implies that
\begin{equation}
\tilde{u}_r B_\theta - \tilde{u}_\theta B_r = f_\eta \overline{\eta} \left[ \frac{1}{r} \frac{\partial}{\partial r}(r B_\phi) - \frac{1}{r} \frac{\partial B_r}{\partial \theta} \right] \mbox{   .   }
\end{equation}
The $\phi-$component of the equation of conservation of magnetic flux:
\begin{eqnarray}
\frac{1}{r}\frac{\partial}{\partial r} \left( r \tilde{u}_\phi B_r - r \tilde{u}_r B_\phi \right) - \frac{1}{r } \frac{\partial }{\partial \theta} \left( \tilde{u}_\theta B_\phi - \tilde{u}_\phi B_\theta \right) \nonumber \\ = - \frac{f_\eta}{r} \frac{\partial }{\partial r} \left[ \overline{\eta} \frac{\partial }{\partial r} (r B_\phi) \right] - \frac{f_\eta \overline{\eta}}{r^2} \frac{\partial}{\partial \theta} \left[ \frac{1}{\st} \frac{\partial}{\partial \theta} ( \st B_\phi) \right]\mbox{   .   }
\end{eqnarray}
Finally, the solenoidal condition is
\begin{equation}
\frac{1}{r^2} \frac{\partial}{\partial r} \left( r^2  B_r \right) + \frac{1}{r \st}\frac{\partial }{\partial \theta} \left (\st B_{\theta} \right) = 0\mbox{   .   }
\end{equation}

\section[]{Numerical implementation of the equations}

Numerical implementation of the equations is done by defining the non-dimensional independent variables $x = r/\rsun$ and $\mu = \ct$, and by working with the non-dimensional dependent variables 
\begin{align}
& \hat{u} = \frac{\tilde{u}_r}{\rsun \Omega_{\rm eq}} \mbox{   ,   } \hat{v} = \frac{\st \tilde{u}_\theta}{\rsun \Omega_{\rm eq}}  \mbox{   ,   } \hat{L} = \frac{r \st \tilde{u}_\phi}{ \rsun^2 \Omega_{\rm eq}} \\ 
& \hat{B} = \frac{B_r}{B_0} \mbox{   ,   } \hat{b} = \frac{\st B_\theta}{B_0}  \mbox{   ,   } \hat{S} = \frac{r \st B_\phi}{\rsun B_0} \mbox{  ,   } \hat{J} =\frac{4\pi r \st j_\phi}{B_0}  \mbox{   .   }  \nonumber 
\end{align}
The perturbed thermodynamical quantities are also normalized with
\begin{equation}
\hat{\rho} = \frac{\tilde{\rho}}{\rho_0}  \mbox{   ,   } \hat{T} = \frac{\tilde{T}}{T_0} \mbox{   ,   } \hat{p} = \frac{\tilde{p}}{p_0}  \mbox{   ,   }
\end{equation}
where $\rho_0 = 1 {\rm g/cm}^3$, $T_0 = 1 {\rm K}$, and $p_0 = \rho_0 \rsun^2 \overline{\Omega}^2$.

The symmetries of the system in the latitudial coordinate $\theta$ suggest the expansion 
of the dependent variables onto Fourier modes, which are equivalent to Chebishev polynomials 
of the variable $\mu$ since the $n-$order Chebishev polynomial $T_n$ is defined as
\begin{equation}
 T_n(\mu) = \cos(n\theta) \mbox{   .  }
\end{equation}
We assume symmetry across the equator for the radial velocity, the angular velocity, the 
thermodynamical quantities and the latitudinal component of the magnetic field. Antisymmetry 
across the equator is then required for consistency for the latitudinal velocity as well as 
the radial and toroidal components of the magnetic field. In addition, we require that the 
total mass flux across a spherical surface be null. This leads to the suggested expansion:
\begin{eqnarray}
\hat{u}(x,\mu) &=& \sum_{n=1}^N \psi_n(x) \frac{\ptl}{\ptl \mu} \left((1-\mu^2) T_{2n-1}(\mu)\right) \nonumber \\
\hat{v}(x,\mu) &=& (1-\mu^2) \sum_{n=1}^N v_n(x) T_{2n-1}(\mu) \mbox{   ,} \nonumber \\
\hat{L}(x,\mu) &=& (1-\mu^2) \sum_{n =1}^N L_n(r) T_{2n-2}(\mu) \mbox{   , }\nonumber  \\
\hat{p}(x,\mu) &=& \sum_{n=1}^N p_n(x) T_{2n-2}(\mu) \mbox{   ,   } \nonumber \\
\hat{\rho}(x,\mu) &=& \sum_{n=1}^N \rho_n(x) T_{2n-2}(\mu) \mbox{   ,} \nonumber \\
\hat{T}(x,\mu) &=& \sum_{n=1}^N \Theta_n(x) T_{2n-2}(\mu) \mbox{   ,}\nonumber 
\end{eqnarray}
\begin{eqnarray}
\hat{B}(x,\mu) &=&  \sum_{n=1}^N B_n(x) T_{2n-1}(\mu) \mbox{   ,   } \nonumber \\
\hat{b}(x,\mu) &=& (1-\mu^2)  \sum_{n=1}^N b_n(x) T_{2n-2}(\mu) \mbox{   ,}\nonumber \\
\hat{S}(x,\mu)  &=& (1-\mu^2)  \sum_{n=1}^N S_n(x) T_{2n-1}(\mu) \mbox{   ,   }\nonumber \\
\hat{J}(x,\mu)  &=& (1-\mu^2)  \sum_{n=1}^N J_n(x) T_{2n-1}(\mu) \mbox{   ,}
\label{eq:defexp}
\end{eqnarray}
where each sum is truncated to retain the first $N$ modes only. 
The equations are expanded using 
the ansatz (\ref{eq:defexp}), then projected back onto the first $N$ 
Chebishev modes (of relevant parity). The quadratic terms are simplified 
analytically using the standard formulae 
$2T_{n}(\mu) T_{m}(\mu) = (T_{n+m}(\mu) + T_{n-m}(\mu))$ 
and $T_n(\mu) = T_{-n}(\mu)$, as well as a variety of others that have been 
summarized by Garaud (2001, pp. 200-204). 
This procedure yields a combination of $12N$ first-order ODEs 
and $3N$ algebraic equations for a total of 
15$N$ dependent variables, namely $\{\psi_n\}$, $\{v_n\}$, 
$\{L_n\}$, $\{T_n\}$ 
and $\{S_n\}$ (and their respective radial first derivatives), as well 
as $\{p_n\}$, $\{\rho_n\}$, $\{B_n\}$, $\{b_n\}$ and $\{J_n\}$ for $n=1..N$. 

These equations are solved using a versatile variant of the 
standard Newton-Raphson-Kantorovich (NRK) relaxation solver (see for example 
Press {\it et al.} 1996, chapter 17.3) developped by Garaud 
(2001, pp. 108-111). The added feature compared with the standard 
algorithm permits the solution of any set of equations that can be 
written in the form
\begin{equation}
\sum_{j = 1}^{N_{\rm v}} M_{ij}({\bf y}, x) \frac{\partial y_j}{\partial x} = f_i({\bf y}, x) \mbox{   for  } i = 1..N_{\rm v} \mbox{  ,  }
\label{eq:mynrk}
\end{equation}
where $x$ is the independent variable, ${\bf y}= \{y_i\}$ is the vector 
containing the $N_{\rm v}$ dependent variables, and $M_{ij}({\bf y}, x)$ 
and $f_i({\bf y}, x)$ can be any nonlinear function of $x$ and ${\bf y}$. 
One of the many advantages of this form is that algebraic equations are 
straightforwardly treated as any other equation by setting $M_{ij} =0$ 
for all $j$ and for $i$ corresponding to the relevant equation(s).

As in the standard relaxation algorithm (see Press {\it et al.} 1996) 
most of the computational cost (both in terms of time and memory)
arises from reading and inverting a block-tridiagonal matrix 
containing $N_x+1$ rows
of 3 blocks\footnote{except for the boundary blocks, of course.} 
(where $N_x$ is equal to the number of meshpoints used, 
typically between 2000 and 3000), 
and where each block has size\footnote{again, except 
for the boundary blocks, which are smaller.}
 $N_{\rm v} \times N_{\rm v}$  where 
$N_{\rm v}$ is the total number of dependent variables ($N_{\rm v} = 15N$, 
where $N$ is 60--80 for a typical run). 
Calculation of the matrix components, followed 
by its inversion using serial partial pivoting can require several hours 
per iteration on a high-end desktop. In addition, memory becomes an issue 
since the standard algorithm typically requires the storage of $N_x$ 
double-precision arrays of size $N_{\rm v} \times N_{\rm v}/2$ before 
back-substitution can proceed. For this reason, we have developped a 
parallel version of the NRK solver.

\section[]{Parallel NRK algorithm}

A serial, simple and efficient way of inverting the typical block-tridiagonal
 linear systems arising from two-point boundary value problems is described 
by Press {\it et al.} (1996). Given the matrix structure
\begin{equation}
\begin{bmatrix}
M & R &   &   &   &   &   &   & \\
L & M & R &   &   &   &   &   & \\
  & L & M & R &   &   &   &   & \\
  &   & L & M & R &   &   &   & \\
  &   &   & L & M & R &   &   & \\
  &   &   &   & L & M & R &   & \\
  &   &   &   &   & L & M & R & \\
  &   &   &   &   &   & L & M & R  \\
  &   &   &   &   &   &   & L & M 
\end{bmatrix}
\end{equation}
the algorithm consists in diagonalizing the first top $M$ (middle) block using 
Gauss-Jordan elimination with partial implicit pivoting, and storing the resulting 
modified $R$ block and right-hand side for later back-substitution. Moving onto the 
second block-row, we zero the $L$ block, diagonalize the $M$ block and store the 
modified $R$ block (and modified right-hand-side vector), and so forth. 
The resulting matrix structure is
\begin{equation}
\begin{bmatrix}
I & S &   &   &   &   &   &   &\\ 0 & I & S &   &   &   &   &   &\\ &
0 & I & S &   &   &   &   &\\ &   & 0 & I & S &   &   &   &\\ &   &
& 0 & I & S &   &   &\\ &   &   &   & 0 & I & S &   &\\ &   &   &   &
& 0 & I & S & \\ &   &   &   &   &   & 0 & I & S\\ &   &   &   &   &
&   & 0 & I
\end{bmatrix}
\end{equation}
where $0$ denotes a ``zeroed'' block,  $I$ is the identity, and $S$ is
a stored block. The last step involves sweeping the matrix back from
bottom to top to back-substitute for the unknown variables. This
method requires minimal storage: firstly, each block-row only needs to
be read just before being processed and secondly, the backsubstitution
step only requires knowledge of the blocks $S$ and modified
right-hand-side vector.

Inverting the same linear system in parallel is not obvious since the
previous algorithm is inherently serial. A standard parallel method
used for  block tridiagonal matrices is cyclic reduction (see Golub \&
Ortega, 1993).  In this case, the number of variables is successively
halved by repeating the  following reduction algorithm: diagonalize
all $M$-blocks in odd block-rows by  partial implicit pivoting, then
zero the $R$ and $L$ matrices directly  above and below by Gaussian
elimination. The resulting matrix after the first reduction is:
\begin{equation}
\begin{bmatrix}
I & S &   &   &   &   &   &   & \\
0 & X & 0 & X &   &   &   &   & \\
  & S & I & S &   &   &   &   & \\
  & X & 0 & X & 0 & X &   &   & \\
  &   &   & S & I & S &   &   & \\
  &   &   & X & 0 & X & 0 & X  & \\
  &   &   &   &   & S & I & S &  \\
  &   &   &   &   & X & 0 & X & 0  \\
  &   &   &   &   &   &   & S & I 
\end{bmatrix}
\end{equation}
where $S$ denotes a block that has been modified and needs to be stored 
for later back-substitution, and $X$ denotes blocks that have been modified 
and will be used in the next reduction step. Note that all of the remaining 
$X$ blocks (in the odd block-rows) form a new block-tridiagonal system 
for roughly half the number of 
variables (more precisely, if the initial number of block-rows is $2^k - 1$ 
then the reduction step reduces it to $2^{k-1}-1$). The algorithm can 
be repeated until only 1 block is left and finally inverted, 
at which point backsubstitution can begin. The main advantage 
of the method is its inherently parallelizable nature, since each process 
can be instructed to reduce a number of blocks more-or-less independently 
of the others; communication between processes is minimal until the final 
steps where the number of remaining block-rows is equal to the number of 
processes. Unfortunately, the stability of this cyclic reduction algorithm 
is much weaker that the stability of the serial algorithm (since the matrix
cannot be globally pivoted), and was found to be unreliable for our purpose. 

An alternative parallel algorithm was constructed, loosely 
based on the work of 
Wright (1992). The block-rows are first equally distributed between 
the processes. A first sweeping reduction akin to the serial algorithm 
is performed starting from the top block-row down to the second-to-last 
block-row in each process\footnote{except in the last process, which is 
swept all the way} to transform the matrix to 
\begin{equation}
\setcounter{MaxMatrixCols}{15}
\begin{bmatrix}
I & S &   &   &   &   &   &   &   &   &   &    \\
0 & I & S &   &   &   &   &   &   &   &   &     \\
  & 0 & I & S &   &   &   &   &   &   &   &       \\
  &   & 0 & X & X &   &   &   &   &   &   &       \\
\hdotsfor{12} \\
  &   &   & S & I & S &   &   &   &   &   &     \\
  &   &   & S & 0 & I & S &   &   &   &   &   \\
  &   &   & S &   & 0 & I & S &   &   &   &     \\
  &   &   & X &   &   & 0 & X & X &   &   &     \\
\hdotsfor{12} \\
  &   &   &   &   &   &   & S & I & S &   &     \\
  &   &   &   &   &   &   & S & 0 & I & S &     \\
  &   &   &   &   &   &   & S &   & 0 & I & S \\
  &   &   &   &   &   &   & X &   &   & 0 & I 
\end{bmatrix}
\end{equation}
(assuming in the case shown here that 12 block-rows are equally 
distributed between 3 processes). Again, by convention all $S$ blocks denote 
blocks that need to be stored for backsubstitution, while $X$ blocks are 
blocks that will be further modified. Until this point, work in each 
process can be done without communication with other processes. 

At the end of the sweeping step, each process sends its very last block-row 
to the following process. Thus the blocks are redistributed as
\begin{equation}
\setcounter{MaxMatrixCols}{15}
\begin{bmatrix}
I & S &   &   &   &   &   &   &   &   &   &    \\
0 & I & S &   &   &   &   &   &   &   &   &     \\
  & 0 & I & S &   &   &   &   &   &   &   &       \\
\hdotsfor{12} \\
  &   & 0 & X & X &   &   &   &   &   &   &       \\
  &   &   & S & I & S &   &   &   &   &   &     \\
  &   &   & S & 0 & I & S &   &   &   &   &   \\
  &   &   & S &   & 0 & I & S &   &   &   &     \\
\hdotsfor{12} \\
  &   &   & X &   &   & 0 & X & X &   &   &     \\
  &   &   &   &   &   &   & S & I & S &   &     \\
  &   &   &   &   &   &   & S & 0 & I & S &     \\
  &   &   &   &   &   &   & S &   & 0 & I & S \\
  &   &   &   &   &   &   & X &   &   & 0 & I 
\end{bmatrix}
\end{equation}
Each process can then continue eliminating undesirable variables from its 
top block-row by Gaussian elimination with the successive block-rows below,
until the following form is achieved:
\begin{equation}
\setcounter{MaxMatrixCols}{15}
\begin{bmatrix}
I & S &   &   &   &   &   &   &   &   &   &    \\
0 & I & S &   &   &   &   &   &   &   &   &     \\
  & 0 & I & S &   &   &   &   &   &   &   &       \\
\hdotsfor{12} \\
  &   & 0 & X & 0 & 0 & 0 & X &   &   &   &       \\
  &   &   & S & I & S &   &   &   &   &   &     \\
  &   &   & S & 0 & I & S &   &   &   &   &   \\
  &   &   & S &   & 0 & I & S &   &   &   &     \\
\hdotsfor{12} \\
  &   &   & X &   &   & 0 & X & 0 & 0 & 0 & X    \\
  &   &   &   &   &   &   & S & I & S &   &     \\
  &   &   &   &   &   &   & S & 0 & I & S &     \\
  &   &   &   &   &   &   & S &   & 0 & I & S \\
  &   &   &   &   &   &   & X &   &   & 0 & I 
\end{bmatrix}
\end{equation}
At this point, the top-rows in each process can be combined into 
a much-reduced block tri-diagonal system. Solution from here on proceeds
with a cyclic reduction of the remaining blocks across processes, 
followed by a back-substitution step.

The operation count of this algorithm is the same as that of 
the standard cyclic reduction. The advantages of this algorithm over
the standard cyclic reduction are two-fold. Firstly, it is found to be much 
more stable (Wright, 1992). Secondly, it is far more versatile, since 
the standard algorithm only performs ideally with a number of block-rows 
equal to $2^k + 1$ (for any integer $k$), which seriously constrains 
the number of meshpoints one is allowed to select. On the other hand, 
in this algorithm it is always possible to balance the load (for any 
number of meshpoints) in such a way that any process contains at most one more
block-row than the average. 

\begin{figure}
\centerline{\epsfig{file=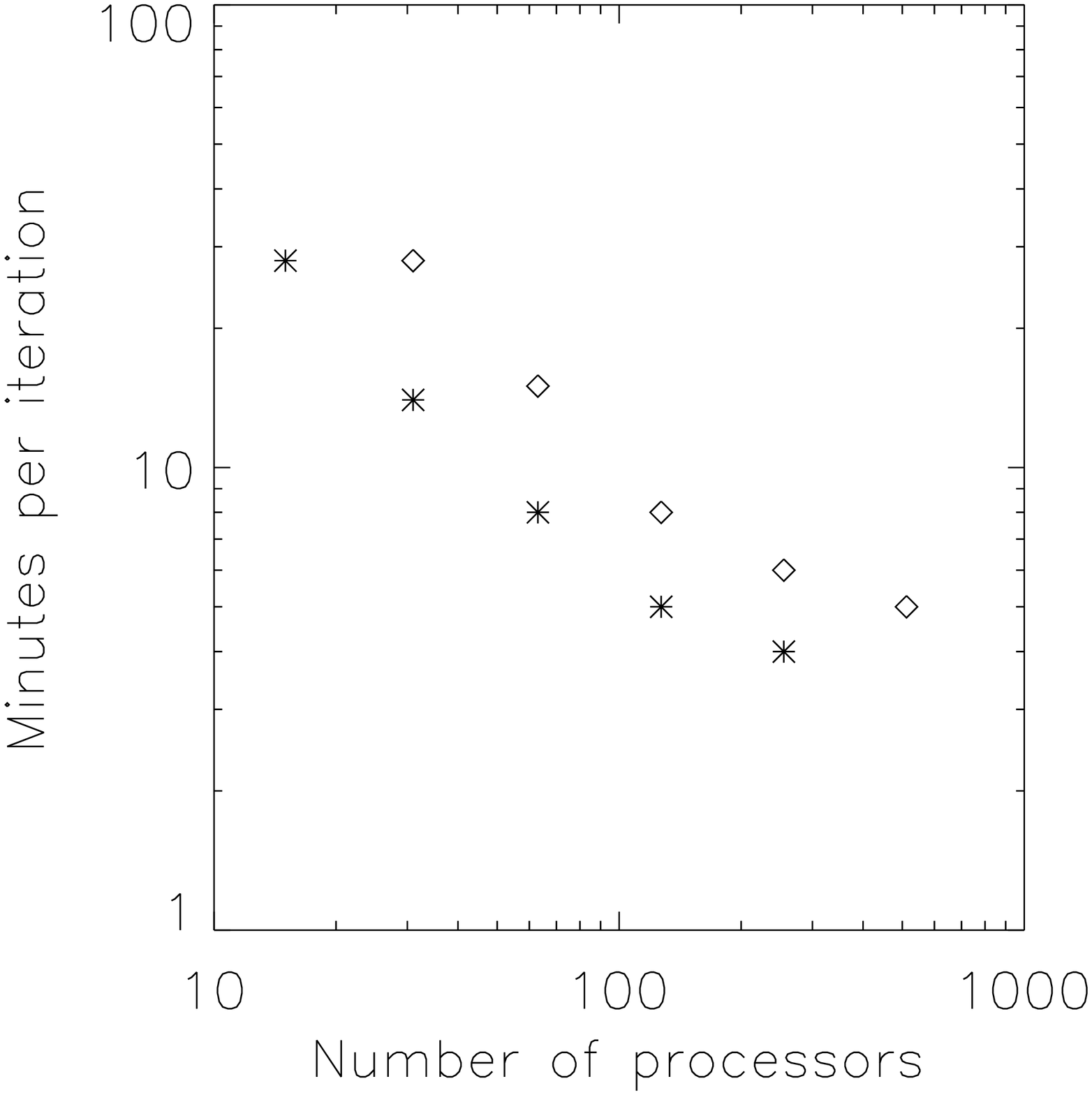,width=8.4cm,height=7cm}}
\caption{Scaling properties of the parallel NRK algorithm RELAX 
for the simulations 
presented in this paper. The stars correspond to a simulation with 60 
modes (i.e. 900 equations), and 1000 meshpoints. The diamonds correspond to 
the same simulation but with 2000 meshpoints.}
\label{fig:procscal}
\end{figure}

The scalability of the algorithm is naturally excellent until the number 
of processes approaches a tenth of the number of meshpoints. 
This is illustrated in Figure \ref{fig:procscal}.

Finally, we note that the user-interface of the complete associated 
parallel version of the Newton-Raphson-Kantorovich algorithm (RELAX) 
is entirely identical to that of the widely used serial NRK algorithm 
written by Gough \& Moore. Transfer between the parallel and serial
version should be transparent to the user. The RELAX algorithm is freely 
available upon request from P. Garaud.

\bsp

\label{lastpage}


\begin{thebibliography}{99}

\bibitem{ah73}
 Acheson, D. J. \& Hide, R., 1973, Rep. Prog. Phys., 36, 159
\bibitem{bal89}
 Brown, T. M., Christensen-Dalsgaard, J. Dziembowsky, W. A., Goode, P., Gough, D. O. \& Morrow, C. A., 1989, ApJ, 343, 526
\bibitem{brunal04} 
 Brun, A.S., Miesch, M.S. \& Toomre, J., 2004, ApJ, 614, 1073
\bibitem{bz06}
 Brun, A. S. \& Zahn, J.-P., 2006, A\&A, 457, 665
\bibitem{jcdschou88} 
 Christensen-Dalsgaard, J. \& Schou, J., 1988,  in {\it Seismology of the Sun and Sun-Like Stars}, ed. V. Domingo \& E.J. Rolfe (ESA-SP286), p. 149
 \bibitem{jcdgoughthompson91}
 Christensen-Dalsgaard, J., Gough D.O. \&  Thompson, M.J., 1991, ApJ, 378, 413
\bibitem{cluneal99} 
 Clune T. L. {\it et al.} 1999, Parallel Comp., 25, 361
\bibitem{dal89}
 Dziembowski, W. A., Goode, P. R. \& Libbrecht, K. G., 1989, ApJ, 337, L53
\bibitem{e25}
 Eddington, A. S., 1925, Observatory, 48,73
\bibitem{eg98}
 Elliott, J. R. \& Gough, D. O., 1999, ApJ, 516, 475
\bibitem{f37}
 Ferraro, V. C. A., 1937, MNRAS, 97, 458
\bibitem{g01}
 Garaud, P., 2001, PhD Thesis, available from \\ http://www.ams.ucsc.edu/$\sim$pgaraud/Work.html
\bibitem{g02}
 Garaud, P., 2002, MNRAS, 329, 1
\bibitem{g07}
 Garaud, P., 2007, in {\it The Solar Tachocline}, pp. 147--181, eds. Hughes, D. W., Rosner, R. \& Weiss, CUP.
\bibitem{gg07}
 Garaud, P. \& Garaud, J.-D., 2007. AMS Technical Report ams2007-17, available from \\ http://www.ams.ucsc.edu/research.html
\bibitem{gb08}
 Garaud, P. \& Brummell, N. H., 2008, ApJ, 674, 498
\bibitem{g89}
 Gilman, P.A., Morrow, C.A. \& Deluca, E.E., 1989, 338, 528
\bibitem{glatzmaier84} 
 Glatzmaier, G.A., 1984, J. Comp. Phys., 55, 461
\bibitem{go93}
 Golub, G. H. \& Ortega, J.M., 1993, in {\it Scientific Computing: An Introduction with Parallel Computing}, pp. 302--321
\bibitem{g85}
 Gough, D. O., 1985, in {\it Proceedings of an ESA Workshop, Garmisch-Parkenkirschen, Germany}
\bibitem{gmi98}
 Gough, D. O. \& McIntyre, M. E., 1998, Nature, 394, 755
\bibitem{go07}
 Gough, D. O., 2007. in {\it The Solar Tachocline}, pp. 3--30, eds. Hughes, D. W., Rosner, R. \& Weiss, CUP.
\bibitem{kr06}
Kitchatinov, L. L. \& R\"udiger, G., 2006, A\&A, 453, 329.
\bibitem{k88}
 Kosovichev, A. G., 1988, Sov. Astron. Lett., 14, 145
\bibitem{miesch00} 
 Miesch, M.S. {\it et al.}, 2000, ApJ, 532, 596.
\bibitem{miesch05}
 Miesch, M.S., 2005, LRSP, 2, 1
\bibitem{mbt06}
 Miesch, M.S., Brun, A.S. \& Toomre, J., 2006, ApJ, 641, 618
\bibitem{mc99}
 MacGregor, K. B. \& Charbonneau, P., 1999, ApJ, 519, 911
\bibitem{m53}
 Mestel, L., 1953, MNRAS, 113, 716
\bibitem{mw87}
 Mestel, L. \& Weiss, N. O., 1987, MNRAS, 226, 123
\bibitem{pal96} 
 Press, W.J. {\it et al.}, 1996, in {\it Numerical Recipes in Fortran 77}, second edition, pp. 753--763.
\bibitem{r05}
 Rempel, M., 2005, ApJ, 622, 1320
\bibitem{rk97}
 R\"{u}diger, G. \& Kitchatinov, L. L., 1997, Astr. Nachr. 318, 273
\bibitem{sal98}
 Schou, J., et al., 1998, ApJ, 505, 390
\bibitem{sz92}
 Spiegel, E. A. \& Zahn, J.-P., 1992, A\&A, 265, 106
\bibitem{sal05}
Sule, A., R\"udiger, G., \& Arlt, R., 2005, A\&A, 437, 1061
\bibitem{s50}
 Sweet, E., 1950, MNRAS, 110, 548
\bibitem{w92}
 Wright, S. J., 1992, SIAM J. Sci. Statist. Comput., 13, 742
\bibitem{z07}
 Zahn, J.-P., 2007, in {\it The Solar Tachocline}, eds. Hughes, D. W., Rosner, R. \& Weiss, CUP.
\end{thebibliography}
\end{document}